\documentclass[aps,pre,reprint,superscriptaddress]{revtex4-1}
\usepackage[dvipdfmx]{graphicx}
\usepackage{dcolumn}
\usepackage{bm}
\usepackage{amssymb}
\usepackage{color}
\begin{document}

\title{\centering{{Laser-Driven Proton-Only Acceleration in a Multicomponent Near-Critical-Density Plasma}}}
\author{Y. Sakawa}%
 \email{sakawa-y@ile.osaka-u.ac.jp}
\affiliation{Institute of Laser Engineering, Osaka University, Suita, Osaka 565-0871, Japan}

\author{H. Ishihara}
\affiliation{Graduate School of Science, Osaka University, Toyonaka, Osaka 560-0043, Japan}

\author{S. N. Ryazantsev}
\affiliation{Joint Institute for High Temperature, RAS, Moscow 125412, Russia}
\affiliation{HB11 Energy Holdings, Freshwater, NSW 2095, Australia}

\author{M. A. Alkhimova}
\affiliation{Joint Institute for High Temperature, RAS, Moscow 125412, Russia}

\author{R. Kumar}
\affiliation{Graduate School of Science, Osaka University, Toyonaka, Osaka 560-0043, Japan}

\author{O. Kuramoto}
\affiliation{Graduate School of Science, Osaka University, Toyonaka, Osaka 560-0043, Japan}

\author{Y. Matsumoto}
\affiliation{Graduate School of Science, Osaka University, Toyonaka, Osaka 560-0043, Japan}

\author{M. Ota}
\affiliation{Graduate School of Science, Osaka University, Toyonaka, Osaka 560-0043, Japan}
\affiliation{National Institute for Fusion Science, Oroshicho, Toki, Gifu 509-5292, Japan}

\author{S. Egashira}
\affiliation{Graduate School of Science, Osaka University, Toyonaka, Osaka 560-0043, Japan}

\author{Y. Nakagawa}
\affiliation{Graduate School of Science, Osaka University, Toyonaka, Osaka 560-0043, Japan}

\author{T. Minami}
\affiliation{Graduate School of Engineering, Osaka University, Suita, Osaka 565-0871, Japan}

\author{K. Sakai}
\affiliation{Graduate School of Engineering, Osaka University, Suita, Osaka 565-0871, Japan}
\affiliation{National Institute for Fusion Science, Oroshicho, Toki, Gifu 509-5292, Japan}

\author{T. Taguchi}
\affiliation{Graduate School of Engineering, Osaka University, Suita, Osaka 565-0871, Japan}

\author{H. Habara}
\affiliation{Graduate School of Engineering, Osaka University, Suita, Osaka 565-0871, Japan}

\author{Y. Kuramitsu}
\affiliation{Graduate School of Engineering, Osaka University, Suita, Osaka 565-0871, Japan}

\author{A. Morace}
\affiliation{Institute of Laser Engineering, Osaka University, Suita, Osaka 565-0871, Japan}

\author{Y. Abe}
\affiliation{Institute of Laser Engineering, Osaka University, Suita, Osaka 565-0871, Japan}

\author{Y. Arikawa}
\affiliation{Institute of Laser Engineering, Osaka University, Suita, Osaka 565-0871, Japan}

\author{S. Fujioka}
\affiliation{Institute of Laser Engineering, Osaka University, Suita, Osaka 565-0871, Japan}

\author{M. Kanasaki}
\affiliation{Graduate School of Maritime Sciences, Kobe University, Kobe 658-0022, Japan}

\author{T. Asai}
\affiliation{Graduate School of Maritime Sciences, Kobe University, Kobe 658-0022, Japan}
\affiliation{Kansai Photon Science Institute (KPSI), National Institutes for Quantum and Radiological Science and Technology (QST), Kizugawa, Kyoto 619-0215, Japan}

\author{T. Morita}
\affiliation{Faculty of Engineering Sciences, Kyushu University, Kasuga, Fukuoka 816-8580, Japan}

\author{Y. Fukuda}
\affiliation{Kansai Photon Science Institute (KPSI), National Institutes for Quantum and Radiological Science and Technology (QST), Kizugawa, Kyoto 619-0215, Japan}

\author{S. Pikuz}
\affiliation{Joint Institute for High Temperature, RAS, Moscow 125412, Russia}
\affiliation{HB11 Energy Holdings, Freshwater, NSW 2095, Australia}

\author{T. Pikuz}
\affiliation{Institute for Open and Transdisciplinary Research Initiatives, Osaka University, Suita, Osaka, 565-0871 Japan}

\author{Y. Ohira}
\affiliation
{Department of Earth and Planetary Science, The University of Tokyo, Bunkyo-ku, Tokyo 113-0033, Japan}

\author{L. N. K. D$\ddot{\rm o}$hl}
\affiliation{York Plasma Institute, School of Physics, Engineering and Technology, University of York, York YO10 5DD, United Kingdom}
\affiliation{Faculty of Biological Sciences, University of Leeds, Leeds LS2 9JT, United Kingdom}

\author{N. Woolsey}
\affiliation{York Plasma Institute, School of Physics, Engineering and Technology, University of York, York YO10 5DD, United Kingdom}

\author{T. Sano}
\affiliation{Institute of Laser Engineering, Osaka University, Suita, Osaka 565-0871, Japan}

\date{\today}
%
%
\begin{abstract}
An experimental investigation of collisionless shock ion acceleration is presented using a multicomponent plasma and a high-intensity picosecond duration laser pulse.
Protons are the only accelerated ions when a near-critical-density plasma is driven by a laser with a modest normalized vector potential.
The results of particle-in-cell simulations imply that collisionless shock may accelerate protons alone selectively,  which can be an important tool for understanding the physics of inaccessible collisionless shocks in space and astrophysical plasma.
\end{abstract}

\maketitle
Collisionless shocks mediated by an ambient magnetic field are ubiquitous in space and astrophysical plasmas and are believed to be sources of high-energy particles or cosmic rays \cite{Sagdeev1966,Bell1978,Blandford1978,Wu1984a,Ball2013,Hoshino2001,Sakawa2016a}.
Unmagnetized electrostatic collisionless shocks \cite{Forslund1970, Forslund1971} are readily created in laboratory plasmas yet are likely rare phenomena in space \cite{Balogh2013} and astrophysical systems.
However, there are common and important collisionless processes resulting from {the reflection and acceleration of particles in the upstream of both magnetized and unmagnetized collisionless shocks, including the excitation of two-stream instabilities \cite{Ohira2008,Treumann2009} and shock dissipation by reflected ions \cite{Treumann2009,Balogh2013,Madanian2020}.
Therefore, understanding collisionless shocks and the associated particle acceleration processes is of general importance in laboratory experiments and astrophysical systems.
Since these processes are difficult to investigate in astrophysical settings due to their remote locations, laboratory experiments are a unique way of studying them.
Laser-driven magnetized collisionless shock experiments have been conducted in the last decade \cite{Woolsey2001,Niemann2014,Schaeffer2016b,Schaeffer2019,Yao2021,Yamazaki2022,Matsukiyo2022}: because of the large externally applied magnetic field, particle reflection from the magnetized collisionless shock experiments are difficult to observe.

A high-intensity laser-driven plasma can generate a strong electrostatic field, accelerating ions to high energies over short distances.
Electrostatic collisionless shock ion acceleration (CSA) \cite{Denavit1992,Silva2004} is one of the laser-driven ion acceleration mechanisms that have been proposed in the last decade \cite{Daido2012,Macchi2013,Bulanov2014a}.
In a laser-driven CSA scheme, the ions located ahead of the shock front are reflected at the shock and accelerated to twice the shock velocity by the electrostatic potential at the shock front.
By studying reflected and accelerated ions from electrostatic collisionless shocks, such experiments may provide access to plasma processes common to collisionless shocks in general.
Furthermore, it might be possible by adjusting the laser and plasma parameters to study the magnetic field generation and amplification mechanisms, such as the Weibel instability \cite{Weibel1959,Fried1959}, the Bell \cite{Bell2004} and the resonant \cite{Kulsrud1969,Wentzel1974} instabilities with a weak external magnetic field, which are predicted to occur in the astrophysical shocks, using the collisionless shock-reflected and accelerated ions.

Various numerical and experimental studies on the CSA mechanism have been carried out \cite{Macchi2012,Haberberger2012,Fiuza2012,Tresca2015,Antici2017,Zhang2017,Chen2017a,Pak2018a,Matsui2019,Ota2019,Singh2020,Tochitsky2020,Boella2021}.
\citet{Haberberger2012} and \citet{Tresca2015} have presented clear experimental evidence for the proton acceleration via the CSA mechanism in a proton plasma using a wavelength of 10-$\mathrm{\mu m}$ laser by measuring a steep electron density profile with the interferometry method when the shock is formed.
In the experiments using 0.8 or $1~\mathrm{\mu m}$ lasers \cite{Antici2017,Zhang2017,Chen2017a,Pak2018a,Ota2019,Singh2020,Tochitsky2020}, on the other hand, the electron density is hardly measured since the density is too high for the interferometry measurement and assumed to be near or below the critical density ($\sim$$10^{21}$ cm$^{-3}$).
Therefore, these works using 0.8- or 1-$\mathrm{\mu m}$ lasers are yet to produce clear signatures demonstrating the generation of a collisionless shock and ion acceleration.

Recently, we demonstrated in two-dimensional (2D) particle-in-cell (PIC) simulations the possibility of producing proton beams in a multicomponent $\rm C_2H_3Cl$ plasma \cite{Kumar2019a,Kumar2021,Sakawa2021}.
\citet{Kumar2021} showed two electrostatic collisionless shocks at two distinct longitudinal positions when driven with laser at normalized laser vector potential $a_0$ $= e E / (m_e \omega c) >5$ (here $e$, $E$, $m_e$, $\omega$, and $c$ are the electric charge, the laser electric field, the electron mass, angular frequency, and the seed of light, respectively).
Moreover, these shocks, associated with protons and carbon (C) ions, accelerate ions to different velocities in an expanding upstream with higher flux than in a single-component hydrogen or C-ion plasma.
On the other hand, when $a_0<5$, a shock forms only in the proton population.
Protons accelerate at this shock, and no C-ion acceleration occurs \cite{Kumar2019a}.
This proton-only acceleration in a multicomponent plasma is an important and new feature of the CSA mechanism but has not been demonstrated in the experiment.
%

In this Letter, we report an experimental investigation of proton-only acceleration in a multicomponent near-critical-density plasma.
A high-intensity ps-pulse laser system with a wavelength of $1~\mathrm{\mu m}$ and a$_0 \simeq 2$ is used as a `drive' laser.
A multicomponent near critical density is formed by pre-ablating a thin foil of $\rm C_8H_7Cl$ using an ns-pulse `ablation' laser system before the irradiation of the drive laser.
The ion energy spectrum is measured simultaneously with several key electron parameters, such as the density, temperature, and energy spectrum.
Without the ablation laser, the measured ion density $n_i$ is close to the solid density $n_{\rm solid}$, and both protons and carbon ions are accelerated.
The proton energy spectrum agrees well with a theoretical model for protons accelerated by the target normal sheath acceleration (TNSA) mechanism \cite{Mora2003} using the measured hot-electron temperature.
When using an ns-duration laser, a near-critical-density plasma is formed by adjusting the timing and energy.
At optimum conditions, only protons are accelerated.
Furthermore, the electron density $n_e$ is measured to ensure that these observations were obtained when driving a near critical density ($n_{\rm cr} = 1.1\times 10^{21} ~\mathrm{cm^{-3}}$ for a wavelength $1~\mathrm{\mu m}$) plasma.
The proton spectrum displays more high-energy protons than expected from the TNSA mechanism.
The 2D PIC simulations support the measured proton-only acceleration in a multicomponent near-critical-density plasma.

The experiments were performed on LFEX laser at the Institute of Laser Engineering, Osaka University.
LFEX delivers up to 210 J/beam of laser energy on target at a wavelength of $1~\mathrm{\mu m}$ in 1.5 ps with one to three beamlets combined.
Spot diameter is about 60 $\mu$m, the incidence angle is $21^\circ$, the nominal intensity is $(3-6) \times10^{18}$ W/cm$^2$, and a$_0$ = 1.6 - 2.1.
To generate a near-critical-density plasma with a long scale length on the rear side of the target, a density structure conducive for CSA, the rear surface of the target is irradiated with an ablation laser 2.5 ns before the drive laser.
This ablation laser uses one of the Gekko XII beams at a wavelength of  $1~\mathrm{\mu m}$ to deliver approximately 3 J across a 700-$\mu$m diameter spot in a 1.3-ns duration pulse.
The nominal laser intensity is $3 \times10^{11}$ W/cm$^2$ and the angle of incidence is $32^\circ$ \cite{supple}.
The target is a foil of ${\rm C_8H_7Cl}$ with a thickness of $1$ $\mu$m.

The plasma density and bulk temperature measurements are conducted with x-ray spectroscopy of highly charged Cl ions using a focusing spectrometer with spatial resolution (FSSR) instrument \cite{Faenov1994}.
FSSR uses a spherically bent crystal ($\alpha$ quartz with Miller indices 101) and provides high spectral resolution ($ \lambda / \Delta \lambda \simeq 3000$ for our experimental conditions). 
FSSR is applied from the rear side of the target and at $42^\circ$ to the laser axis.
Thomson parabola spectrometer (TPS) \cite{Kuramitsu2022} and electron spectrometer (ESM) located on the axis and $21^\circ$ from the drive laser beam, respectively, (or $21^\circ$ and  $42^\circ$, respectively, from the target normal direction) at the rear side, are used to measure the ion and electron spectra from the rear side of the target.
Image plates (IP), BAS-TR2025/Fuji Film, detect FSSR, TPS, and ESM. 

Figure \ref{fig:Spectra_with_without} shows the experimental (shaded areas) x-ray spectra of highly charged Cl ions using FSSR without [Fig.~\ref{fig:Spectra_with_without}(a)] and with [Fig.~\ref{fig:Spectra_with_without}(b)] the ablation laser. 
The spectra cover the wavelength of  3.5 - 4.5 $\rm{\AA}$.
These spectra are normalized to the emission intensity of the Cl Ly$_\alpha$ line (4.185 $\rm{\AA}$).
Solid lines in Fig.~\ref{fig:Spectra_with_without} represent the modeled x-ray spectra using the collisional-radiative spectral analysis code PrismSPECT \cite{PrismSPECT,MacFarlane2004}. 
 %
  \begin{figure}[]
  \includegraphics[width=\linewidth]{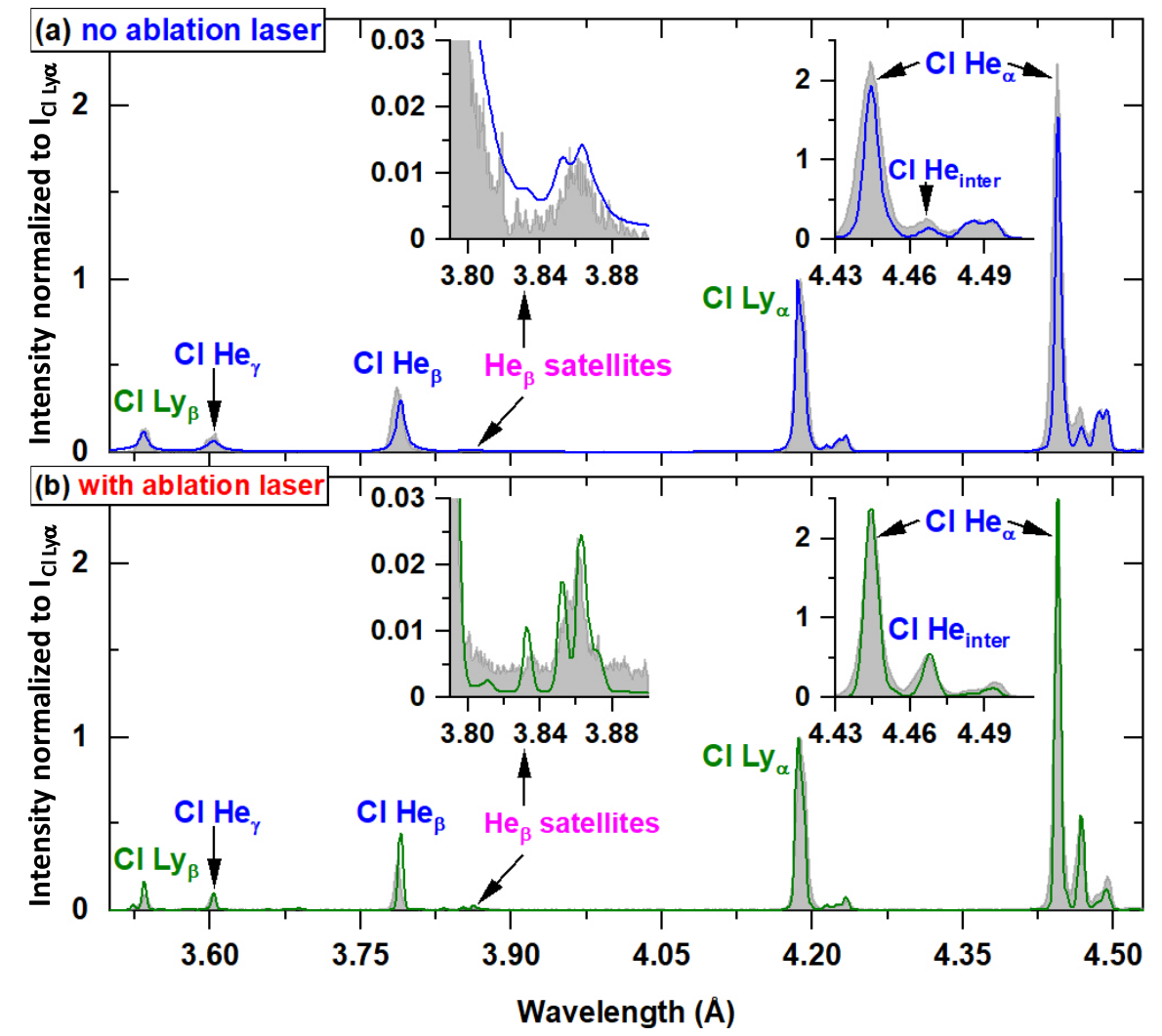}
  \caption{Comparison of the experimental (shaded areas) and modeled (solid lines) x-ray spectra of highly charged Cl ions from the rear side of ${\rm C_8H_7Cl}$ target  (a) without and (b) with the ablation laser. These spectra are normalized to the emission intensity of the Cl Ly$\alpha$ line (4.185 $\rm{\AA}$). 
Insets are expanded spectra of He$_\beta$ satellites, He$_\alpha$, and He$_{\rm{inter}}$.
 Matching modeling uses the collisional-radiative spectral analysis code PrismSPECT \cite{PrismSPECT,MacFarlane2004}. $a_0$ is 1.6 and 2.1 for (a) and (b), respectively.}
   \label{fig:Spectra_with_without}
\end{figure}
For the case without the ablation laser [Fig.~\ref{fig:Spectra_with_without}(a)], the best fit was obtained, assuming that the plasma consists of two zones with different temperatures.
The first zone, which represents a hot zone, is determined by the intensities of the lines emitted by H-like ions.
First, the intensity ratio of Ly$_\alpha$ and dielectronic satellite lines Ly$_{\alpha\rm{-sat}}$$/$Ly$_\alpha$ is used to adjust the electron temperature $T_e$.
Furthermore, to reproduce the shape of Ly$_\alpha$, the effect of self-absorption is included by assuming that the plasma is a sphere with a radius $R$.
Finally,  the width of all the observed lines is reproduced by adjusting $n_i$.
As a result,  plasma parameters of the hot zone are determined as $n_i = (3.2 \pm 0.4) \times 10^{22}$ cm$^{-3}$, $T_e = 1300 \pm 50$ eV, and $R=1.0 \pm 0.2$ $\mu$m.
The second zone, a cold zone, is determined by the intensities of He-like ions, which the hot zone parameters cannot fit.
The intensity ratio of He$_\alpha$ (4.44 $\rm{\AA}$) and intercombination lines He$_{\rm{inter}}$$/$He$_\alpha$ is used to adjust $n_i$ and $T_e$, and results in $n_i = (3.2 \pm 0.4) \times 10^{22}$ cm$^{-3}$, $T_e = 600 \pm 25$ eV, and $R=1.0 \pm 0.2$ $\mu$m.
We see that the experimental and modeled x-ray spectra agree with each other.
The calculated average charge is $\langle Z \rangle = 4.3$ and the electron density is $n_e = (1.4^{+0.1}_{-0.2}) \times 10^{23}$ cm$^{-3}$.
Note that the measured $n_i = (3.2 \pm 0.4) \times 10^{22}$ cm$^{-3}$ is close to the solid density $n_{\rm solid} = 7.6 \times 10^{22}$ cm$^{-3}$ of ${\rm C_8H_7Cl}$ target.

For the case with the ablation laser [Fig.~\ref{fig:Spectra_with_without}(b)], nearly 2 orders of magnitude smaller $n_i$ are derived from a significantly higher intensity ratio of He$_{\rm{inter}}$$/$He$_\alpha$.
In this case, the ns ablation laser produces a uniform pre-plasma with small temperature and density gradients. Accordingly, the plasma driven by the ps drive laser is also homogeneous, and a one-zone model is sufficient to describe the results.
The intensity ratios of He$_{\rm{inter}}$ and He$_\alpha$, and their intensities relative to that of Ly$_\alpha$ are used to adjust  $n_i$ and $T_e$.
The self-absorption of the plasma is also included.
As a result, $n_i = (3.0 \pm 0.5) \times 10^{20}$ cm$^{-3}$, and $T_e = 1370 \pm 60$ eV are obtained.
In the range of physically reasonable values of $R$ = 0 - 8 $\mu$m, the shape of the theoretical spectrum at fixed $n_i$ value depends on weakly on $R$.  Therefore, we do not give the exact value of $R$, but the theoretical spectrum shown in Fig.~\ref{fig:Spectra_with_without}(b) is obtained for $R = 0.25$ $\mu$m.
The modeled x-ray spectrum is shown by the solid line in Fig.~\ref{fig:Spectra_with_without}(b) and agrees well with that obtained from the experiment.
The calculated average charge $\langle Z \rangle = 4.3$ and the electron density $n_e = (1.3 \pm 0.2) \times 10^{21}$ cm$^{-3}$.
The measured $n_e$ is close to the critical density $n_{\rm cr} = 1.1 \times 10^{21}$ cm$^{-3}$, and we confirmed that a suitable plasma condition for CSA is achieved when using an ablation laser.

%
  \begin{figure}[]
  \includegraphics[width=\linewidth]{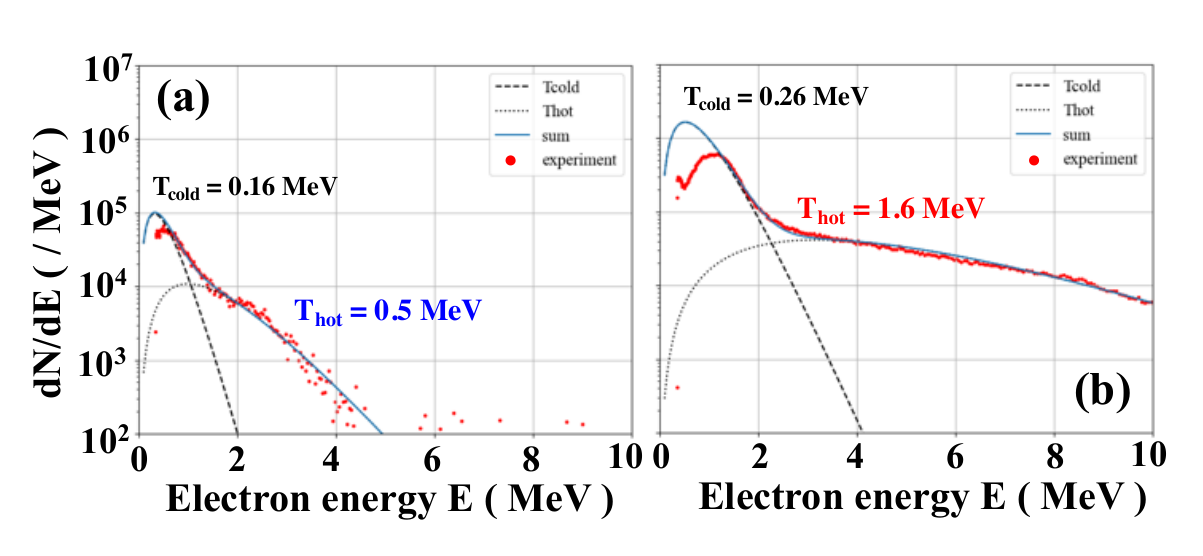}
  \caption{Derived energy spectra of electrons (red dots) (a) without and (b) with the ablation laser. Two-temperature 3D relativistic Maxwellian distribution [Eq.~(\ref{eq:Te})] is used to fit the measured electron energy spectra. The cold and hot components are displayed with the dotted and broken lines, respectively, and the sum of the two Maxwellian fits is represented by the solid line. The derived temperatures are (a) $T_{\rm cold} = 0.16$ and $T_{\rm hot}=0.5$ MeV without the ablation laser, and (b) $T_{\rm cold} = 0.26$ and $T_{\rm hot}=1.6$ MeV with the ablation laser. $a_0$ is 1.6 and 2.0 for (a) and (b), respectively.}
   \label{fig:ESM_w_wo}
\end{figure}

Figure \ref{fig:ESM_w_wo} shows the electron energy spectrum (red dots) obtained from ESM  without [Fig.~\ref{fig:ESM_w_wo}(a)] and with [Fig.~\ref{fig:ESM_w_wo}(b)] the ablation laser.
To derive the electron temperature, we use the following two-temperature 3D relativistic Maxwellian distribution,
%
\begin{equation}
\frac{dN}{dE} = c_{\rm cold} E^2 {\rm exp}\left( -\frac{E}{T_{\rm cold}}\right) + c_{\rm hot} E^2 {\rm exp}\left( -\frac{E}{T_{\rm hot}}\right) .
\label{eq:Te}
\end{equation}
Here, $N$ and $E$ are, respectively, the number and energy of the electrons, $c_{\rm cold}$ ($c_{\rm hot}$) and $T_{\rm cold}$ ($T_{\rm hot}$) are a scaling factor and the electron temperature for the cold (hot) component, respectively.
In Fig.~\ref{fig:ESM_w_wo}, fitted functions of the cold and hot components are displayed with dotted and broken lines, respectively, and the sum of the two Maxwellian fits is represented by a solid line.
The two-temperature approximation reproduces the measured electron spectrum in the $E > 1$ MeV energy range.
The shifts in the peak energy of the cold components result from the plasma potential at the target's rear side.
The $T_e$ derived from FSSR is roughly 2 orders smaller than $T_{\rm cold}$.
This is because FSSR measures the temperature of electrons within the bulk plasma, while the ESM measures electrons with energies sufficient to leave the plasma.
%
  \begin{figure}[]
  \includegraphics[width=\linewidth]{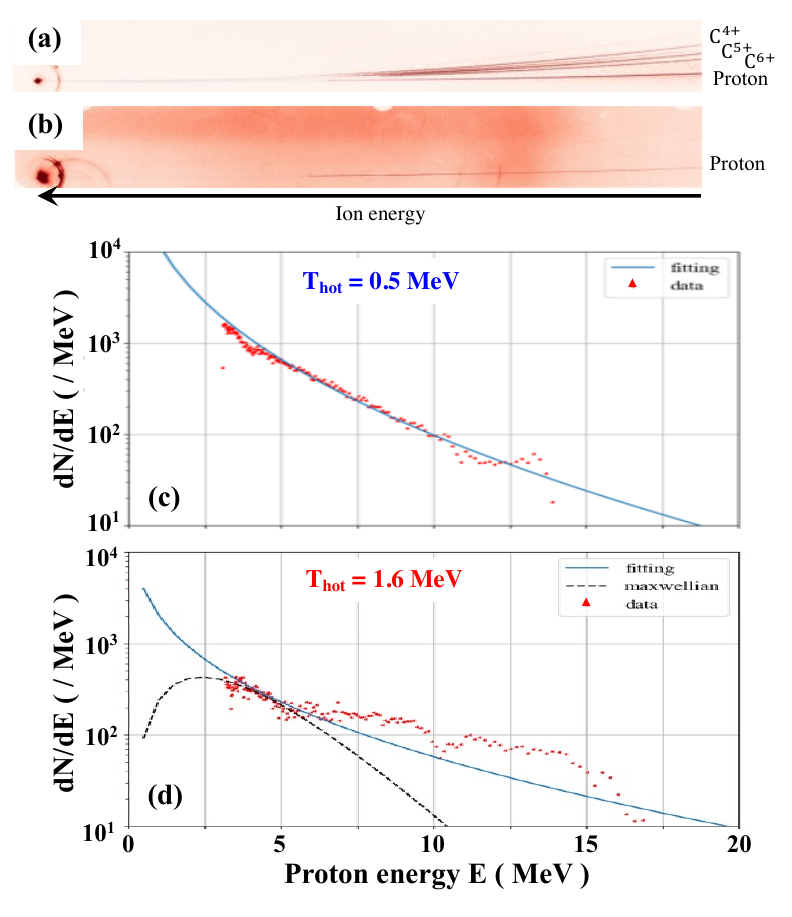}
    \caption{Raw data of TPS  (a) without and (b) with the ablation laser. $a_0$ is 1.6 and 2.0 for (a) and (b), respectively. Derived energy spectra of protons (red dots) (c) without and (d) with the ablation laser.
The error bars or standard deviations in the $y$-axis in (d) are less than 1\% in the energy ranges of $E = 6 - 15$ MeV. Theoretical energy spectra of protons accelerated by the TNSA mechanism [Eq.~(\ref{eq:TNSA})] (solid lines) are shown in (c) and (d) for the measured hot-electron temperatures of $T_{\rm hot}=0.5$ and 1.6 MeV, respectively.
The dashed line in (d) fits the Maxwellian proton distribution.}
   \label{fig:TP_w_wo}
\end{figure}

Figures \ref{fig:TP_w_wo}(a) and  \ref{fig:TP_w_wo}(b) display raw data of TPS without and with the ablation laser, respectively.
We fit the parabola signals with the least squares fitting method \cite{Kuramitsu2022}.
An infinite energy point is determined from the dot on the left side of the IP.
Background signals are taken from the trajectory at 10 pixels from each signal.
Signals from protons and carbon ions appear when the drive laser alone is used, but only a proton signal is detected when the ablation laser is used.

Figures \ref{fig:TP_w_wo}(c) and \ref{fig:TP_w_wo}(d) present inferred energy spectra of protons (red dots) without and with the ablation laser, respectively.
\citet{Mora2003} has derived a theoretical energy spectrum for the ions accelerated by the TNSA mechanism using a hot-electron temperature ($T_{\rm hot}$) as follows,
%
\begin{equation}
\frac{dN}{dE} = \frac{c_{\rm ion}}{\sqrt{E {T_{\rm hot}}}} {\rm exp} \left(-\sqrt{\frac{E}{T_{\rm hot}}}\right),
\label{eq:TNSA}
\end{equation}
where $N$ and $E$ are the number and energy of the protons, and $c_{\rm ion}$ is a scaling factor.
We adjust $c_{\rm ion}$ so that Eq.~(\ref{eq:TNSA}) fits the measured spectra.
For the no-ablation-laser case [Fig.~\ref{fig:TP_w_wo}(c)], the measured ion spectrum is in good agreement with the TNSA model (the solid line) with the measured electron temperature $T_{\rm hot} = 0.5$ MeV.
On the other hand, with the ablation laser [Fig.~\ref{fig:TP_w_wo}(d)], the theoretical energy spectrum of the TNSA protons using the measured $T_{\rm hot}=1.6$ MeV (the solid line) fits the measured spectrum only around the proton energy of $E$ = 2.6 - 6.0 MeV, and above $E \simeq 6.0$ MeV, the number of the accelerated protons is larger than the TNSA model.
The dashed line in Fig.~\ref{fig:TP_w_wo}(d) is the Maxwellian distribution.
These results suggest that when the ablation laser is used and the plasma density is near critical, some mechanisms other than TNSA  contribute to ion acceleration.

To identify the mechanism of the proton-only acceleration, we examine numerical simulations of CSA in a multicomponent $\rm C_8H_7Cl$ plasma, using the 2D PIC simulation code EPOCH \cite{Arber2015}.
The laser parameters are $a_0 = 2.1$ and a Gaussian temporal profile with $1.5 ~\mathrm{ps}$ full-width-at-half-maximum, and the maximum electron density is the relativistic critical density $a_0 n_{cr}$\cite{Kumar2019a,Kumar2021,Sakawa2021}.
Details of the simulations, including the target density profiles, are given in \citet{Kumar2019a}. 

 \begin{figure}
  \includegraphics[width=\linewidth]{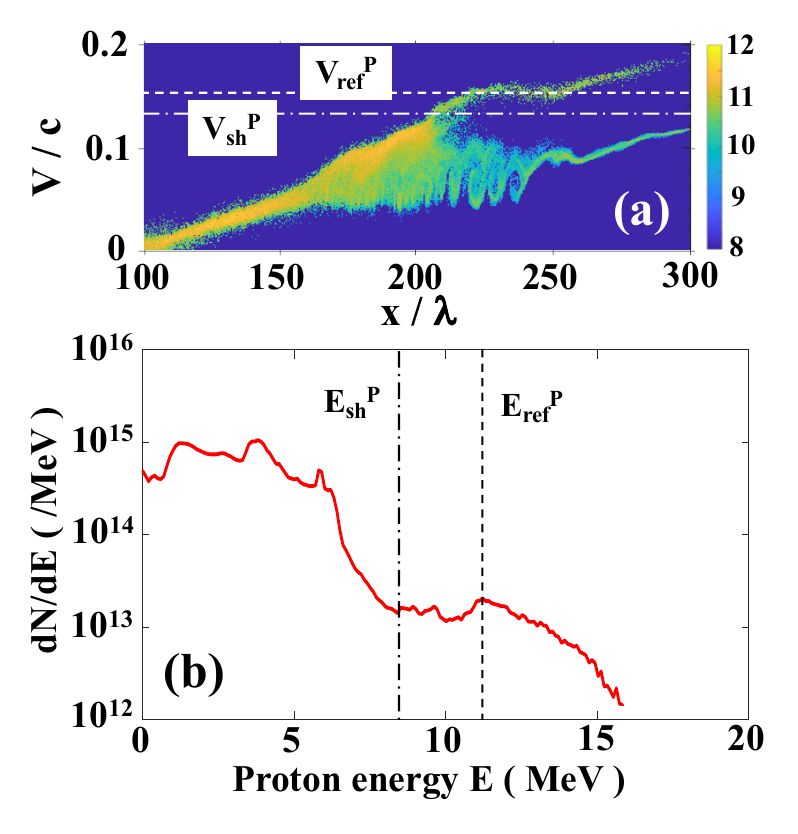}
   \caption{2D PIC simulation results for a  $\rm C_8H_7Cl$ plasma at $a_0 = 2.1$ and \textit{t} = 6.75 ps. (a) Phase space of protons and (b) energy spectra of the protons taken with $v_x > 0$. 
   The color scale shows the number of protons on a log scale.
   The velocities (energies) of the shock $V_{sh}^{\rm P}$ ($E_{\rm sh}^{\rm P}$) and the reflected protons $V_{ref}^{\rm P}$ ($E_{ref}^{\rm P}$) are shown in the dash-dotted and dotted lines, respectively.}
  \label{C8H7Cl_spect}
\end{figure}

Figures \ref{C8H7Cl_spect}(a) and \ref{C8H7Cl_spect}(b) represent the proton phase space and the energy spectrum taken with $v_x > 0$, respectively.
The velocities (energies) of the shock $V_{sh}^{\rm P}$ ($E_{sh}^{\rm P}$) and the reflected protons $V_{ref}^{\rm P}$ ($E_{ref}^{\rm P}$) are shown in the dash-dotted and dotted lines, respectively.
A significant population of protons satisfies the condition for CSA, $v_L^i \leq v_0^i \leq V_{sh}^i$ \cite{Tidman,Kumar2019a,Kumar2021,Sakawa2021}, where $v_{0}^i$ is the drift velocity of the upstream ions, $v_L^i = V_{sh}^i - \sqrt{2(Z_i /A_i)} e\phi / m_p$ is the lower-threshold velocity, $Z_i$ and $A_i$ are the ionic charge and mass numbers, respectively, $e$ is the electric charge, $\phi$ is the electrostatic potential at the shock front, $m_p$ is the proton mass, the higher-threshold velocity $V^i_{sh}$ is the shock velocity, and $i$ represents the different ion species ($i=\rm P$ and C for protons and C$^{6+}$ ions, respectively).
As a result, protons in the upstream region are reflected and accelerated at the collisionless shock \cite{Kumar2019a,Kumar2021,Sakawa2021}. 
However, no C$^{6+}$ ions are reflected by this collisionless shock, as this requires $V_{sh}^{\rm C}$ = $V_{sh}^{\rm P}$ since the velocities of upstream C$^{6+}$ ions are lower than $v_L^{\rm C}$.
Furthermore, the carbon-ion Mach number \cite{Kumar2021} $M^{\rm C} <1 $ for $a_0 = 2.1$, and no shock is associated with C$^{6+}$ ions.
Furthermore, the maximum proton energies obtained from the experiment [Fig.~\ref{fig:TP_w_wo}(d)] and PIC [Fig.~\ref{C8H7Cl_spect}(b)] agree very well with each other. 
Therefore, proton-only acceleration may occur via the CSA mechanism.

\citet{Fuchs2007} have measured the TNSA-proton energy spectra as a function of the scale lengths of the rear-side plasma ($\l_{\rm rear}$) by changing the timing of the second high-intensity laser-beam irradiation on the rear side of the target from the drive laser beam.
Without the second laser beam or when $\l_{\rm rear}=0$, the maximum proton energy was $\simeq$13.5 MeV.
When $\l_{\rm rear}=9.2$ $\mu$m, the maximum proton energy decreased to $\simeq$7.5 MeV.
Note that the proton spectrum up to the proton energy of $\simeq$6 MeV was nearly identical for both cases.
Therefore, in their experiment, the maximum density is close to solid density. 
Protons are accelerated by TNSA, and the maximum proton energy decreases when the second laser is used because of a reduced rear-side sheath electric field \cite{Fuchs2007}.
Our case differs from \citet{Fuchs2007}, as the ablation laser duration is long and the $n_e \simeq n_{\rm cr}$.
We observe a number of protons larger than that predicted for TNSA.
As calculated by the 2D PIC simulation, CSA can accelerate only protons at $a_0 \simeq 2$.
We suggest this experiment isolates the CSA and that this is the only mechanism accelerating ions, and in this case, only protons.
{We have used a multicomponent (protons and C$^{6+}$ ions) in a near-critical-density plasma at $a_0 \simeq 2$.

In summary, we investigated collisionless shock ion acceleration (CSA) using a high-intensity ps-pulse laser system with a modest normalized vector potential.
In a near-critical-density plasma containing the multicomponent ion species C$_8$H$_7$Cl, only protons are accelerated.

In a near-critical-density multicomponent proton and C$^{6+}$-ion plasma, 2D PIC calculations suggest that it might be possible to control accelerated ions actively (proton-only or proton and C$^{6+}$-ion acceleration) by changing the drive-laser intensity in CSA.
The collisionless shock-reflected and accelerated ions are essential in the Galactic cosmic ray generation via diffusive shock acceleration \cite{Caprioli2017,Caprioli2015,Ohira2016a}, and the magnetic field generation and amplification via instabilities \cite{Weibel1959,Fried1959,Bell2004,Kulsrud1969,Wentzel1974}.
This work illustrates how laboratory studies of ion acceleration at collisionless shocks can be an important tool for understanding some collisionless physics associated with space and astrophysical shocks.
%
%

This research was partially supported by the Japan Society for the Promotion of Science (JSPS) KAKENHI Grant No. JP15H02154, JP17H06202, JP19H00668, JP19H01893, JP22H00119, JSPS Core-to-Core Program B. Asia-Africa Science Platforms Grant No. JPJSCCB20190003, and UK EPSRC grants EP/L01663X/1 and EP/P026796/1. 
%
%


\begin{thebibliography}{54}%
\makeatletter
\providecommand \@ifxundefined [1]{%
 \@ifx{#1\undefined}
}%
\providecommand \@ifnum [1]{%
 \ifnum #1\expandafter \@firstoftwo
 \else \expandafter \@secondoftwo
 \fi
}%
\providecommand \@ifx [1]{%
 \ifx #1\expandafter \@firstoftwo
 \else \expandafter \@secondoftwo
 \fi
}%
\providecommand \natexlab [1]{#1}%
\providecommand \enquote  [1]{``#1''}%
\providecommand \bibnamefont  [1]{#1}%
\providecommand \bibfnamefont [1]{#1}%
\providecommand \citenamefont [1]{#1}%
\providecommand \href@noop [0]{\@secondoftwo}%
\providecommand \href [0]{\begingroup \@sanitize@url \@href}%
\providecommand \@href[1]{\@@startlink{#1}\@@href}%
\providecommand \@@href[1]{\endgroup#1\@@endlink}%
\providecommand \@sanitize@url [0]{\catcode `\\12\catcode `\$12\catcode
  `\&12\catcode `\#12\catcode `\^12\catcode `\_12\catcode `\%12\relax}%
\providecommand \@@startlink[1]{}%
\providecommand \@@endlink[0]{}%
\providecommand \url  [0]{\begingroup\@sanitize@url \@url }%
\providecommand \@url [1]{\endgroup\@href {#1}{\urlprefix }}%
\providecommand \urlprefix  [0]{URL }%
\providecommand \Eprint [0]{\href }%
\providecommand \doibase [0]{http://dx.doi.org/}%
\providecommand \selectlanguage [0]{\@gobble}%
\providecommand \bibinfo  [0]{\@secondoftwo}%
\providecommand \bibfield  [0]{\@secondoftwo}%
\providecommand \translation [1]{[#1]}%
\providecommand \BibitemOpen [0]{}%
\providecommand \bibitemStop [0]{}%
\providecommand \bibitemNoStop [0]{.\EOS\space}%
\providecommand \EOS [0]{\spacefactor3000\relax}%
\providecommand \BibitemShut  [1]{\csname bibitem#1\endcsname}%
\let\auto@bib@innerbib\@empty
\bibitem [{\citenamefont {Sagdeev}(1966)}]{Sagdeev1966}%
  \BibitemOpen
  \bibfield  {author} {\bibinfo {author} {\bibfnamefont {R.~Z.}\ \bibnamefont
  {Sagdeev}},\ }\href {http://adsabs.harvard.edu/abs/1966RvPP....4...23S}
  {\emph {\bibinfo {title} {Reviews of Plasma Physics}}},\ Vol.~\bibinfo
  {volume} {4}\ (\bibinfo {year} {1966})\ pp.\ \bibinfo {pages}
  {23--91}\BibitemShut {NoStop}%
  \bibitem [{\citenamefont {Bell}(1978)}]{Bell1978}%
  \BibitemOpen
  \bibfield  {author} {\bibinfo {author} {\bibfnamefont {A.~R.}\ \bibnamefont
  {Bell}},\ }\href {\doibase 10.1093/mnras/182.2.147} {\bibfield  {journal}
  {\bibinfo  {journal} {Monthly Notices of the Royal Astronomical Society}\
  }\textbf {\bibinfo {volume} {182}},\ \bibinfo {pages} {147} (\bibinfo {year}
  {1978})}\BibitemShut {NoStop}%
\bibitem [{\citenamefont {Blandford}\ and\ \citenamefont
  {Ostriker}(1978)}]{Blandford1978}%
  \BibitemOpen
  \bibfield  {author} {\bibinfo {author} {\bibfnamefont {R.~D.}\ \bibnamefont
  {Blandford}}\ and\ \bibinfo {author} {\bibfnamefont {J.~P.}\ \bibnamefont
  {Ostriker}},\ }\href {\doibase 10.1086/182658} {\bibfield  {journal}
  {\bibinfo  {journal} {The Astrophysical Journal}\ }\textbf {\bibinfo {volume}
  {221}},\ \bibinfo {pages} {L29} (\bibinfo {year} {1978})}\BibitemShut
  {NoStop}%
\bibitem [{\citenamefont {Wu}(1984)}]{Wu1984a}%
  \BibitemOpen
  \bibfield  {author} {\bibinfo {author} {\bibfnamefont {C.~S.}\ \bibnamefont
  {Wu}},\ }\href {\doibase 10.1029/JA089iA10p08857} {\bibfield  {journal}
  {\bibinfo  {journal} {Journal of Geophysical Research}\ }\textbf {\bibinfo
  {volume} {89}},\ \bibinfo {pages} {8857} (\bibinfo {year}
  {1984})}\BibitemShut {NoStop}%
\bibitem [{\citenamefont {Ball}\ and\ \citenamefont
  {Melrose}(2001)}]{Ball2013}%
  \BibitemOpen
  \bibfield  {author} {\bibinfo {author} {\bibfnamefont {L.}~\bibnamefont
  {Ball}}\ and\ \bibinfo {author} {\bibfnamefont {D.~B.}\ \bibnamefont
  {Melrose}},\ }\href {\doibase 10.1071/AS01047} {\bibfield  {journal}
  {\bibinfo  {journal} {Publications of the Astronomical Society of Australia}\
  }\textbf {\bibinfo {volume} {18}},\ \bibinfo {pages} {361} (\bibinfo {year}
  {2001})}\BibitemShut {NoStop}%
\bibitem [{\citenamefont {Hoshino}(2001)}]{Hoshino2001}%
  \BibitemOpen
  \bibfield  {author} {\bibinfo {author} {\bibfnamefont {M.}~\bibnamefont
  {Hoshino}},\ }\href {\doibase 10.1143/PTPS.143.149} {\bibfield  {journal}
  {\bibinfo  {journal} {Progress of Theoretical Physics Supplement}\ }\textbf
  {\bibinfo {volume} {143}},\ \bibinfo {pages} {149} (\bibinfo {year}
  {2001})}\BibitemShut {NoStop}%
\bibitem [{\citenamefont {Sakawa}\ \emph {et~al.}(2016)\citenamefont {Sakawa},
  \citenamefont {Morita}, \citenamefont {Kuramitsu},\ and\ \citenamefont
  {Takabe}}]{Sakawa2016a}%
  \BibitemOpen
  \bibfield  {author} {\bibinfo {author} {\bibfnamefont {Y.}~\bibnamefont
  {Sakawa}}, \bibinfo {author} {\bibfnamefont {T.}~\bibnamefont {Morita}},
  \bibinfo {author} {\bibfnamefont {Y.}~\bibnamefont {Kuramitsu}}, \ and\
  \bibinfo {author} {\bibfnamefont {H.}~\bibnamefont {Takabe}},\ }\href
  {\doibase 10.1080/23746149.2016.1213660} {\bibfield  {journal} {\bibinfo
  {journal} {Advances in Physics: X}\ }\textbf {\bibinfo {volume} {1}},\
  \bibinfo {pages} {425} (\bibinfo {year} {2016})}\BibitemShut {NoStop}%
\bibitem [{\citenamefont {Forslund}\ and\ \citenamefont
  {Shonk}(1970)}]{Forslund1970}%
  \BibitemOpen
  \bibfield  {author} {\bibinfo {author} {\bibfnamefont {D.}~\bibnamefont
  {Forslund}}\ and\ \bibinfo {author} {\bibfnamefont {C.}~\bibnamefont
  {Shonk}},\ }\href {\doibase 10.1103/PhysRevLett.25.281} {\bibfield  {journal}
  {\bibinfo  {journal} {Physical Review Letters}\ }\textbf {\bibinfo {volume}
  {25}},\ \bibinfo {pages} {281} (\bibinfo {year} {1970})}\BibitemShut
  {NoStop}%
\bibitem [{\citenamefont {Forslund}\ \emph {et~al.}(1971)\citenamefont
  {Forslund}, \citenamefont {Morse},\ and\ \citenamefont
  {Nielson}}]{Forslund1971}%
  \BibitemOpen
  \bibfield  {author} {\bibinfo {author} {\bibfnamefont {D.~W.}\ \bibnamefont
  {Forslund}}, \bibinfo {author} {\bibfnamefont {R.~L.}\ \bibnamefont {Morse}},
  \ and\ \bibinfo {author} {\bibfnamefont {C.~W.}\ \bibnamefont {Nielson}},\
  }\href {\doibase 10.1103/PhysRevLett.27.1424} {\bibfield  {journal} {\bibinfo
   {journal} {Physical Review Letters}\ }\textbf {\bibinfo {volume} {27}},\
  \bibinfo {pages} {1424} (\bibinfo {year} {1971})}\BibitemShut {NoStop}%
\bibitem [{\citenamefont {Balogh}\ and\ \citenamefont
  {Treumann}(2013)}]{Balogh2013}%
  \BibitemOpen
  \bibfield  {author} {\bibinfo {author} {\bibfnamefont {A.}~\bibnamefont
  {Balogh}}\ and\ \bibinfo {author} {\bibfnamefont {R.~A.}\ \bibnamefont
  {Treumann}},\ }\href@noop {} {\emph {\bibinfo {title} {{Physics of
  Collisionless Shocks: Space Plasma Shock Waves}}}}\ (\bibinfo  {publisher}
  {Springer Science {\&} Business Media, New York},\ \bibinfo {year}
  {2013})\BibitemShut {NoStop}%
\bibitem [{\citenamefont {Ohira}\ and\ \citenamefont
  {Takahara}(2008)}]{Ohira2008}%
  \BibitemOpen
  \bibfield  {author} {\bibinfo {author} {\bibfnamefont {Y.}~\bibnamefont
  {Ohira}}\ and\ \bibinfo {author} {\bibfnamefont {F.}~\bibnamefont
  {Takahara}},\ }\href {\doibase 10.1086/592182} {\bibfield  {journal}
  {\bibinfo  {journal} {The Astrophysical Journal}\ }\textbf {\bibinfo {volume}
  {688}},\ \bibinfo {pages} {320} (\bibinfo {year} {2008})},\ \Eprint
  {http://arxiv.org/abs/0808.3195} {arXiv:0808.3195} \BibitemShut {NoStop}%
\bibitem [{\citenamefont {Treumann}(2009)}]{Treumann2009}%
  \BibitemOpen
  \bibfield  {author} {\bibinfo {author} {\bibfnamefont {R.~A.}\ \bibnamefont
  {Treumann}},\ }\href {\doibase 10.1007/s00159-009-0024-2} {\bibfield
  {journal} {\bibinfo  {journal} {The Astronomy and Astrophysics Review}\
  }\textbf {\bibinfo {volume} {17}},\ \bibinfo {pages} {409} (\bibinfo {year}
  {2009})}\BibitemShut {NoStop}%
\bibitem [{\citenamefont {Madanian}\ \emph {et~al.}(2020)\citenamefont
  {Madanian}, \citenamefont {Schwartz}, \citenamefont {Halekas},\ and\
  \citenamefont {Wilson}}]{Madanian2020}%
  \BibitemOpen
  \bibfield  {author} {\bibinfo {author} {\bibfnamefont {H.}~\bibnamefont
  {Madanian}}, \bibinfo {author} {\bibfnamefont {S.~J.}\ \bibnamefont
  {Schwartz}}, \bibinfo {author} {\bibfnamefont {J.~S.}\ \bibnamefont
  {Halekas}}, \ and\ \bibinfo {author} {\bibfnamefont {L.~B.}\ \bibnamefont
  {Wilson}},\ }\href {\doibase 10.1029/2020GL088309} {\bibfield  {journal}
  {\bibinfo  {journal} {Geophysical Research Letters}\ }\textbf {\bibinfo
  {volume} {47}},\ \bibinfo {pages} {1} (\bibinfo {year} {2020})}\BibitemShut
  {NoStop}%
\bibitem [{\citenamefont {Woolsey}\ \emph {et~al.}(2001)\citenamefont
  {Woolsey}, \citenamefont {Ali}, \citenamefont {Evans}, \citenamefont
  {Grundy}, \citenamefont {Pestehe}, \citenamefont {Carolan}, \citenamefont
  {Conway}, \citenamefont {Dendy}, \citenamefont {Helander}, \citenamefont
  {McClements}, \citenamefont {Kirk}, \citenamefont {Norreys}, \citenamefont
  {Notley},\ and\ \citenamefont {Rose}}]{Woolsey2001}%
  \BibitemOpen
  \bibfield  {author} {\bibinfo {author} {\bibfnamefont {N.~C.}\ \bibnamefont
  {Woolsey}}, \bibinfo {author} {\bibfnamefont {Y.~A.}\ \bibnamefont {Ali}},
  \bibinfo {author} {\bibfnamefont {R.~G.}\ \bibnamefont {Evans}}, \bibinfo
  {author} {\bibfnamefont {R.~a.~D.}\ \bibnamefont {Grundy}}, \bibinfo {author}
  {\bibfnamefont {S.~J.}\ \bibnamefont {Pestehe}}, \bibinfo {author}
  {\bibfnamefont {P.~G.}\ \bibnamefont {Carolan}}, \bibinfo {author}
  {\bibfnamefont {N.~J.}\ \bibnamefont {Conway}}, \bibinfo {author}
  {\bibfnamefont {R.~O.}\ \bibnamefont {Dendy}}, \bibinfo {author}
  {\bibfnamefont {P.}~\bibnamefont {Helander}}, \bibinfo {author}
  {\bibfnamefont {K.~G.}\ \bibnamefont {McClements}}, \bibinfo {author}
  {\bibfnamefont {J.~G.}\ \bibnamefont {Kirk}}, \bibinfo {author}
  {\bibfnamefont {P.~a.}\ \bibnamefont {Norreys}}, \bibinfo {author}
  {\bibfnamefont {M.~M.}\ \bibnamefont {Notley}}, \ and\ \bibinfo {author}
  {\bibfnamefont {S.~J.}\ \bibnamefont {Rose}},\ }\href {\doibase
  10.1063/1.1351831} {\bibfield  {journal} {\bibinfo  {journal} {Physics of
  Plasmas}\ }\textbf {\bibinfo {volume} {8}},\ \bibinfo {pages} {2439}
  (\bibinfo {year} {2001})}\BibitemShut {NoStop}%
\bibitem [{\citenamefont {Niemann}\ \emph {et~al.}(2014)\citenamefont
  {Niemann}, \citenamefont {Gekelman}, \citenamefont {Constantin},
  \citenamefont {Everson}, \citenamefont {Schaeffer}, \citenamefont
  {Bondarenko}, \citenamefont {Clark}, \citenamefont {Winske}, \citenamefont
  {Vincena}, \citenamefont {{Van Compernolle}},\ and\ \citenamefont
  {Pribyl}}]{Niemann2014}%
  \BibitemOpen
  \bibfield  {author} {\bibinfo {author} {\bibfnamefont {C.}~\bibnamefont
  {Niemann}}, \bibinfo {author} {\bibfnamefont {W.}~\bibnamefont {Gekelman}},
  \bibinfo {author} {\bibfnamefont {C.~G.}\ \bibnamefont {Constantin}},
  \bibinfo {author} {\bibfnamefont {E.~T.}\ \bibnamefont {Everson}}, \bibinfo
  {author} {\bibfnamefont {D.~B.}\ \bibnamefont {Schaeffer}}, \bibinfo {author}
  {\bibfnamefont {A.~S.}\ \bibnamefont {Bondarenko}}, \bibinfo {author}
  {\bibfnamefont {S.~E.}\ \bibnamefont {Clark}}, \bibinfo {author}
  {\bibfnamefont {D.}~\bibnamefont {Winske}}, \bibinfo {author} {\bibfnamefont
  {S.}~\bibnamefont {Vincena}}, \bibinfo {author} {\bibfnamefont
  {B.}~\bibnamefont {{Van Compernolle}}}, \ and\ \bibinfo {author}
  {\bibfnamefont {P.}~\bibnamefont {Pribyl}},\ }\href {\doibase
  10.1002/2014GL061820} {\bibfield  {journal} {\bibinfo  {journal} {Geophysical
  Research Letters}\ }\textbf {\bibinfo {volume} {41}},\ \bibinfo {pages}
  {7413} (\bibinfo {year} {2014})}\BibitemShut {NoStop}%
\bibitem [{\citenamefont {Schaeffer}\ \emph {et~al.}(2017)\citenamefont
  {Schaeffer}, \citenamefont {Fox}, \citenamefont {Haberberger}, \citenamefont
  {Fiksel}, \citenamefont {Bhattacharjee}, \citenamefont {Barnak},
  \citenamefont {Hu},\ and\ \citenamefont {Germaschewski}}]{Schaeffer2016b}%
  \BibitemOpen
  \bibfield  {author} {\bibinfo {author} {\bibfnamefont {D.~B.}\ \bibnamefont
  {Schaeffer}}, \bibinfo {author} {\bibfnamefont {W.}~\bibnamefont {Fox}},
  \bibinfo {author} {\bibfnamefont {D.}~\bibnamefont {Haberberger}}, \bibinfo
  {author} {\bibfnamefont {G.}~\bibnamefont {Fiksel}}, \bibinfo {author}
  {\bibfnamefont {A.}~\bibnamefont {Bhattacharjee}}, \bibinfo {author}
  {\bibfnamefont {D.~H.}\ \bibnamefont {Barnak}}, \bibinfo {author}
  {\bibfnamefont {S.~X.}\ \bibnamefont {Hu}}, \ and\ \bibinfo {author}
  {\bibfnamefont {K.}~\bibnamefont {Germaschewski}},\ }\href {\doibase
  10.1103/PhysRevLett.119.025001} {\bibfield  {journal} {\bibinfo  {journal}
  {Physical Review Letters}\ }\textbf {\bibinfo {volume} {119}},\ \bibinfo
  {pages} {025001} (\bibinfo {year} {2017})},\ \Eprint
  {http://arxiv.org/abs/1610.06533} {arXiv:1610.06533} \BibitemShut {NoStop}%
\bibitem [{\citenamefont {Schaeffer}\ \emph {et~al.}(2019)\citenamefont
  {Schaeffer}, \citenamefont {Fox}, \citenamefont {Follett}, \citenamefont
  {Fiksel}, \citenamefont {Li}, \citenamefont {Matteucci}, \citenamefont
  {Bhattacharjee},\ and\ \citenamefont {Germaschewski}}]{Schaeffer2019}%
  \BibitemOpen
  \bibfield  {author} {\bibinfo {author} {\bibfnamefont {D.~B.}\ \bibnamefont
  {Schaeffer}}, \bibinfo {author} {\bibfnamefont {W.}~\bibnamefont {Fox}},
  \bibinfo {author} {\bibfnamefont {R.~K.}\ \bibnamefont {Follett}}, \bibinfo
  {author} {\bibfnamefont {G.}~\bibnamefont {Fiksel}}, \bibinfo {author}
  {\bibfnamefont {C.~K.}\ \bibnamefont {Li}}, \bibinfo {author} {\bibfnamefont
  {J.}~\bibnamefont {Matteucci}}, \bibinfo {author} {\bibfnamefont
  {A.}~\bibnamefont {Bhattacharjee}}, \ and\ \bibinfo {author} {\bibfnamefont
  {K.}~\bibnamefont {Germaschewski}},\ }\href {\doibase
  10.1103/PhysRevLett.122.245001} {\bibfield  {journal} {\bibinfo  {journal}
  {Physical Review Letters}\ }\textbf {\bibinfo {volume} {122}},\ \bibinfo
  {pages} {245001} (\bibinfo {year} {2019})}\BibitemShut {NoStop}%
\bibitem [{\citenamefont {Yao}\ \emph {et~al.}(2021)\citenamefont {Yao},
  \citenamefont {Fazzini}, \citenamefont {Chen}, \citenamefont {Burdonov},
  \citenamefont {Antici}, \citenamefont {B{\'{e}}ard}, \citenamefont
  {Bola{\~{n}}os}, \citenamefont {Ciardi}, \citenamefont {Diab}, \citenamefont
  {Filippov}, \citenamefont {Kisyov}, \citenamefont {Lelasseux}, \citenamefont
  {Miceli}, \citenamefont {Moreno}, \citenamefont {Nastasa}, \citenamefont
  {Orlando}, \citenamefont {Pikuz}, \citenamefont {Popescu}, \citenamefont
  {Revet}, \citenamefont {Ribeyre}, \citenamefont {D'Humi{\`{e}}res},\ and\
  \citenamefont {Fuchs}}]{Yao2021}%
  \BibitemOpen
  \bibfield  {author} {\bibinfo {author} {\bibfnamefont {W.}~\bibnamefont
  {Yao}}, \bibinfo {author} {\bibfnamefont {A.}~\bibnamefont {Fazzini}},
  \bibinfo {author} {\bibfnamefont {S.~N.}\ \bibnamefont {Chen}}, \bibinfo
  {author} {\bibfnamefont {K.}~\bibnamefont {Burdonov}}, \bibinfo {author}
  {\bibfnamefont {P.}~\bibnamefont {Antici}}, \bibinfo {author} {\bibfnamefont
  {J.}~\bibnamefont {B{\'{e}}ard}}, \bibinfo {author} {\bibfnamefont
  {S.}~\bibnamefont {Bola{\~{n}}os}}, \bibinfo {author} {\bibfnamefont
  {A.}~\bibnamefont {Ciardi}}, \bibinfo {author} {\bibfnamefont
  {R.}~\bibnamefont {Diab}}, \bibinfo {author} {\bibfnamefont {E.~D.}\
  \bibnamefont {Filippov}}, \bibinfo {author} {\bibfnamefont {S.}~\bibnamefont
  {Kisyov}}, \bibinfo {author} {\bibfnamefont {V.}~\bibnamefont {Lelasseux}},
  \bibinfo {author} {\bibfnamefont {M.}~\bibnamefont {Miceli}}, \bibinfo
  {author} {\bibfnamefont {Q.}~\bibnamefont {Moreno}}, \bibinfo {author}
  {\bibfnamefont {V.}~\bibnamefont {Nastasa}}, \bibinfo {author} {\bibfnamefont
  {S.}~\bibnamefont {Orlando}}, \bibinfo {author} {\bibfnamefont
  {S.}~\bibnamefont {Pikuz}}, \bibinfo {author} {\bibfnamefont {D.~C.}\
  \bibnamefont {Popescu}}, \bibinfo {author} {\bibfnamefont {G.}~\bibnamefont
  {Revet}}, \bibinfo {author} {\bibfnamefont {X.}~\bibnamefont {Ribeyre}},
  \bibinfo {author} {\bibfnamefont {E.}~\bibnamefont {D'Humi{\`{e}}res}}, \
  and\ \bibinfo {author} {\bibfnamefont {J.}~\bibnamefont {Fuchs}},\ }\href
  {\doibase 10.1038/s41567-021-01325-w} {\bibfield  {journal} {\bibinfo
  {journal} {Nature Physics}\ }\textbf {\bibinfo {volume} {17}},\ \bibinfo
  {pages} {1177} (\bibinfo {year} {2021})},\ \Eprint
  {http://arxiv.org/abs/2011.00135} {arXiv:2011.00135} \BibitemShut {NoStop}%
\bibitem [{\citenamefont {Yamazaki}\ \emph {et~al.}(2022)\citenamefont
  {Yamazaki}, \citenamefont {Matsukiyo}, \citenamefont {Morita}, \citenamefont
  {Tanaka}, \citenamefont {Umeda}, \citenamefont {Aihara}, \citenamefont
  {Edamoto}, \citenamefont {Egashira}, \citenamefont {Hatsuyama}, \citenamefont
  {Higuchi}, \citenamefont {Hihara}, \citenamefont {Horie}, \citenamefont
  {Hoshino}, \citenamefont {Ishii}, \citenamefont {Ishizaka}, \citenamefont
  {Itadani}, \citenamefont {Izumi}, \citenamefont {Kambayashi}, \citenamefont
  {Kakuchi}, \citenamefont {Katsuki}, \citenamefont {Kawamura}, \citenamefont
  {Kawamura}, \citenamefont {Kisaka}, \citenamefont {Kojima}, \citenamefont
  {Konuma}, \citenamefont {Kumar}, \citenamefont {Minami}, \citenamefont
  {Miyata}, \citenamefont {Moritaka}, \citenamefont {Murakami}, \citenamefont
  {Nagashima}, \citenamefont {Nakagawa}, \citenamefont {Nishimoto},
  \citenamefont {Nishioka}, \citenamefont {Ohira}, \citenamefont {Ohnishi},
  \citenamefont {Ota}, \citenamefont {Ozaki}, \citenamefont {Sano},
  \citenamefont {Sakai}, \citenamefont {Sei}, \citenamefont {Shiota},
  \citenamefont {Shoji}, \citenamefont {Sugiyama}, \citenamefont {Suzuki},
  \citenamefont {Takagi}, \citenamefont {Toda}, \citenamefont {Tomita},
  \citenamefont {Tomiya}, \citenamefont {Yoneda}, \citenamefont {Takezaki},
  \citenamefont {Tomita}, \citenamefont {Kuramitsu},\ and\ \citenamefont
  {Sakawa}}]{Yamazaki2022}%
  \BibitemOpen
  \bibfield  {author} {\bibinfo {author} {\bibfnamefont {R.}~\bibnamefont
  {Yamazaki}}, \bibinfo {author} {\bibfnamefont {S.}~\bibnamefont {Matsukiyo}},
  \bibinfo {author} {\bibfnamefont {T.}~\bibnamefont {Morita}}, \bibinfo
  {author} {\bibfnamefont {S.~J.}\ \bibnamefont {Tanaka}}, \bibinfo {author}
  {\bibfnamefont {T.}~\bibnamefont {Umeda}}, \bibinfo {author} {\bibfnamefont
  {K.}~\bibnamefont {Aihara}}, \bibinfo {author} {\bibfnamefont
  {M.}~\bibnamefont {Edamoto}}, \bibinfo {author} {\bibfnamefont
  {S.}~\bibnamefont {Egashira}}, \bibinfo {author} {\bibfnamefont
  {R.}~\bibnamefont {Hatsuyama}}, \bibinfo {author} {\bibfnamefont
  {T.}~\bibnamefont {Higuchi}}, \bibinfo {author} {\bibfnamefont
  {T.}~\bibnamefont {Hihara}}, \bibinfo {author} {\bibfnamefont
  {Y.}~\bibnamefont {Horie}}, \bibinfo {author} {\bibfnamefont
  {M.}~\bibnamefont {Hoshino}}, \bibinfo {author} {\bibfnamefont
  {A.}~\bibnamefont {Ishii}}, \bibinfo {author} {\bibfnamefont
  {N.}~\bibnamefont {Ishizaka}}, \bibinfo {author} {\bibfnamefont
  {Y.}~\bibnamefont {Itadani}}, \bibinfo {author} {\bibfnamefont
  {T.}~\bibnamefont {Izumi}}, \bibinfo {author} {\bibfnamefont
  {S.}~\bibnamefont {Kambayashi}}, \bibinfo {author} {\bibfnamefont
  {S.}~\bibnamefont {Kakuchi}}, \bibinfo {author} {\bibfnamefont
  {N.}~\bibnamefont {Katsuki}}, \bibinfo {author} {\bibfnamefont
  {R.}~\bibnamefont {Kawamura}}, \bibinfo {author} {\bibfnamefont
  {Y.}~\bibnamefont {Kawamura}}, \bibinfo {author} {\bibfnamefont
  {S.}~\bibnamefont {Kisaka}}, \bibinfo {author} {\bibfnamefont
  {T.}~\bibnamefont {Kojima}}, \bibinfo {author} {\bibfnamefont
  {A.}~\bibnamefont {Konuma}}, \bibinfo {author} {\bibfnamefont
  {R.}~\bibnamefont {Kumar}}, \bibinfo {author} {\bibfnamefont
  {T.}~\bibnamefont {Minami}}, \bibinfo {author} {\bibfnamefont
  {I.}~\bibnamefont {Miyata}}, \bibinfo {author} {\bibfnamefont
  {T.}~\bibnamefont {Moritaka}}, \bibinfo {author} {\bibfnamefont
  {Y.}~\bibnamefont {Murakami}}, \bibinfo {author} {\bibfnamefont
  {K.}~\bibnamefont {Nagashima}}, \bibinfo {author} {\bibfnamefont
  {Y.}~\bibnamefont {Nakagawa}}, \bibinfo {author} {\bibfnamefont
  {T.}~\bibnamefont {Nishimoto}}, \bibinfo {author} {\bibfnamefont
  {Y.}~\bibnamefont {Nishioka}}, \bibinfo {author} {\bibfnamefont
  {Y.}~\bibnamefont {Ohira}}, \bibinfo {author} {\bibfnamefont
  {N.}~\bibnamefont {Ohnishi}}, \bibinfo {author} {\bibfnamefont
  {M.}~\bibnamefont {Ota}}, \bibinfo {author} {\bibfnamefont {N.}~\bibnamefont
  {Ozaki}}, \bibinfo {author} {\bibfnamefont {T.}~\bibnamefont {Sano}},
  \bibinfo {author} {\bibfnamefont {K.}~\bibnamefont {Sakai}}, \bibinfo
  {author} {\bibfnamefont {S.}~\bibnamefont {Sei}}, \bibinfo {author}
  {\bibfnamefont {J.}~\bibnamefont {Shiota}}, \bibinfo {author} {\bibfnamefont
  {Y.}~\bibnamefont {Shoji}}, \bibinfo {author} {\bibfnamefont
  {K.}~\bibnamefont {Sugiyama}}, \bibinfo {author} {\bibfnamefont
  {D.}~\bibnamefont {Suzuki}}, \bibinfo {author} {\bibfnamefont
  {M.}~\bibnamefont {Takagi}}, \bibinfo {author} {\bibfnamefont
  {H.}~\bibnamefont {Toda}}, \bibinfo {author} {\bibfnamefont {S.}~\bibnamefont
  {Tomita}}, \bibinfo {author} {\bibfnamefont {S.}~\bibnamefont {Tomiya}},
  \bibinfo {author} {\bibfnamefont {H.}~\bibnamefont {Yoneda}}, \bibinfo
  {author} {\bibfnamefont {T.}~\bibnamefont {Takezaki}}, \bibinfo {author}
  {\bibfnamefont {K.}~\bibnamefont {Tomita}}, \bibinfo {author} {\bibfnamefont
  {Y.}~\bibnamefont {Kuramitsu}}, \ and\ \bibinfo {author} {\bibfnamefont
  {Y.}~\bibnamefont {Sakawa}},\ }\href {\doibase 10.1103/physreve.105.025203}
  {\bibfield  {journal} {\bibinfo  {journal} {Physical Review E}\ }\textbf
  {\bibinfo {volume} {105}},\ \bibinfo {pages} {025203} (\bibinfo {year}
  {2022})},\ \Eprint {http://arxiv.org/abs/2201.07976} {arXiv:2201.07976}
  \BibitemShut {NoStop}%
\bibitem [{\citenamefont {Matsukiyo}\ \emph {et~al.}(2022)\citenamefont
  {Matsukiyo}, \citenamefont {Yamazaki}, \citenamefont {Morita}, \citenamefont
  {Tomita}, \citenamefont {Kuramitsu}, \citenamefont {Sano}, \citenamefont
  {Tanaka}, \citenamefont {Takezaki}, \citenamefont {Isayama}, \citenamefont
  {Higuchi}, \citenamefont {Murakami}, \citenamefont {Horie}, \citenamefont
  {Katsuki}, \citenamefont {Hatsuyama}, \citenamefont {Edamoto}, \citenamefont
  {Nishioka}, \citenamefont {Takagi}, \citenamefont {Kojima}, \citenamefont
  {Tomita}, \citenamefont {Ishizaka}, \citenamefont {Kakuchi}, \citenamefont
  {Sei}, \citenamefont {Sugiyama}, \citenamefont {Aihara}, \citenamefont
  {Kambayashi}, \citenamefont {Ota}, \citenamefont {Egashira}, \citenamefont
  {Izumi}, \citenamefont {Minami}, \citenamefont {Nakagawa}, \citenamefont
  {Sakai}, \citenamefont {Iwamoto}, \citenamefont {Ozaki},\ and\ \citenamefont
  {Sakawa}}]{Matsukiyo2022}%
  \BibitemOpen
  \bibfield  {author} {\bibinfo {author} {\bibfnamefont {S.}~\bibnamefont
  {Matsukiyo}}, \bibinfo {author} {\bibfnamefont {R.}~\bibnamefont {Yamazaki}},
  \bibinfo {author} {\bibfnamefont {T.}~\bibnamefont {Morita}}, \bibinfo
  {author} {\bibfnamefont {K.}~\bibnamefont {Tomita}}, \bibinfo {author}
  {\bibfnamefont {Y.}~\bibnamefont {Kuramitsu}}, \bibinfo {author}
  {\bibfnamefont {T.}~\bibnamefont {Sano}}, \bibinfo {author} {\bibfnamefont
  {S.~J.}\ \bibnamefont {Tanaka}}, \bibinfo {author} {\bibfnamefont
  {T.}~\bibnamefont {Takezaki}}, \bibinfo {author} {\bibfnamefont
  {S.}~\bibnamefont {Isayama}}, \bibinfo {author} {\bibfnamefont
  {T.}~\bibnamefont {Higuchi}}, \bibinfo {author} {\bibfnamefont
  {H.}~\bibnamefont {Murakami}}, \bibinfo {author} {\bibfnamefont
  {Y.}~\bibnamefont {Horie}}, \bibinfo {author} {\bibfnamefont
  {N.}~\bibnamefont {Katsuki}}, \bibinfo {author} {\bibfnamefont
  {R.}~\bibnamefont {Hatsuyama}}, \bibinfo {author} {\bibfnamefont
  {M.}~\bibnamefont {Edamoto}}, \bibinfo {author} {\bibfnamefont
  {H.}~\bibnamefont {Nishioka}}, \bibinfo {author} {\bibfnamefont
  {M.}~\bibnamefont {Takagi}}, \bibinfo {author} {\bibfnamefont
  {T.}~\bibnamefont {Kojima}}, \bibinfo {author} {\bibfnamefont
  {S.}~\bibnamefont {Tomita}}, \bibinfo {author} {\bibfnamefont
  {N.}~\bibnamefont {Ishizaka}}, \bibinfo {author} {\bibfnamefont
  {S.}~\bibnamefont {Kakuchi}}, \bibinfo {author} {\bibfnamefont
  {S.}~\bibnamefont {Sei}}, \bibinfo {author} {\bibfnamefont {K.}~\bibnamefont
  {Sugiyama}}, \bibinfo {author} {\bibfnamefont {K.}~\bibnamefont {Aihara}},
  \bibinfo {author} {\bibfnamefont {S.}~\bibnamefont {Kambayashi}}, \bibinfo
  {author} {\bibfnamefont {M.}~\bibnamefont {Ota}}, \bibinfo {author}
  {\bibfnamefont {S.}~\bibnamefont {Egashira}}, \bibinfo {author}
  {\bibfnamefont {T.}~\bibnamefont {Izumi}}, \bibinfo {author} {\bibfnamefont
  {T.}~\bibnamefont {Minami}}, \bibinfo {author} {\bibfnamefont
  {Y.}~\bibnamefont {Nakagawa}}, \bibinfo {author} {\bibfnamefont
  {K.}~\bibnamefont {Sakai}}, \bibinfo {author} {\bibfnamefont
  {M.}~\bibnamefont {Iwamoto}}, \bibinfo {author} {\bibfnamefont
  {N.}~\bibnamefont {Ozaki}}, \ and\ \bibinfo {author} {\bibfnamefont
  {Y.}~\bibnamefont {Sakawa}},\ }\href {\doibase 10.1103/physreve.106.025205}
  {\bibfield  {journal} {\bibinfo  {journal} {Physical Review E}\ }\textbf
  {\bibinfo {volume} {106}},\ \bibinfo {pages} {025205} (\bibinfo {year}
  {2022})},\ \Eprint {http://arxiv.org/abs/2207.12586} {arXiv:2207.12586}
  \BibitemShut {NoStop}%
\bibitem [{\citenamefont {Denavit}(1992)}]{Denavit1992}%
  \BibitemOpen
  \bibfield  {author} {\bibinfo {author} {\bibfnamefont {J.}~\bibnamefont
  {Denavit}},\ }\href {\doibase 10.1103/PhysRevLett.69.3052} {\bibfield
  {journal} {\bibinfo  {journal} {Physical Review Letters}\ }\textbf {\bibinfo
  {volume} {69}},\ \bibinfo {pages} {3052} (\bibinfo {year}
  {1992})}\BibitemShut {NoStop}%
\bibitem [{\citenamefont {Silva}\ \emph {et~al.}(2004)\citenamefont {Silva},
  \citenamefont {Marti}, \citenamefont {Davies}, \citenamefont {Fonseca},
  \citenamefont {Ren}, \citenamefont {Tsung},\ and\ \citenamefont
  {Mori}}]{Silva2004}%
  \BibitemOpen
  \bibfield  {author} {\bibinfo {author} {\bibfnamefont {L.~O.}\ \bibnamefont
  {Silva}}, \bibinfo {author} {\bibfnamefont {M.}~\bibnamefont {Marti}},
  \bibinfo {author} {\bibfnamefont {J.~R.}\ \bibnamefont {Davies}}, \bibinfo
  {author} {\bibfnamefont {R.~A.}\ \bibnamefont {Fonseca}}, \bibinfo {author}
  {\bibfnamefont {C.}~\bibnamefont {Ren}}, \bibinfo {author} {\bibfnamefont
  {F.}~\bibnamefont {Tsung}}, \ and\ \bibinfo {author} {\bibfnamefont {W.~B.}\
  \bibnamefont {Mori}},\ }\href {\doibase 10.1103/PhysRevLett.92.015002}
  {\bibfield  {journal} {\bibinfo  {journal} {Physical Review Letters}\
  }\textbf {\bibinfo {volume} {92}},\ \bibinfo {pages} {015002} (\bibinfo
  {year} {2004})}\BibitemShut {NoStop}%
\bibitem [{\citenamefont {Daido}\ \emph {et~al.}(2012)\citenamefont {Daido},
  \citenamefont {Nishiuchi},\ and\ \citenamefont {Pirozhkov}}]{Daido2012}%
  \BibitemOpen
  \bibfield  {author} {\bibinfo {author} {\bibfnamefont {H.}~\bibnamefont
  {Daido}}, \bibinfo {author} {\bibfnamefont {M.}~\bibnamefont {Nishiuchi}}, \
  and\ \bibinfo {author} {\bibfnamefont {A.~S.}\ \bibnamefont {Pirozhkov}},\
  }\href {\doibase 10.1088/0034-4885/75/5/056401} {\bibfield  {journal}
  {\bibinfo  {journal} {Reports on progress in physics. Physical Society (Great
  Britain)}\ }\textbf {\bibinfo {volume} {75}},\ \bibinfo {pages} {056401}
  (\bibinfo {year} {2012})}\BibitemShut {NoStop}%
\bibitem [{\citenamefont {Macchi}\ \emph {et~al.}(2013)\citenamefont {Macchi},
  \citenamefont {Borghesi},\ and\ \citenamefont {Passoni}}]{Macchi2013}%
  \BibitemOpen
  \bibfield  {author} {\bibinfo {author} {\bibfnamefont {A.}~\bibnamefont
  {Macchi}}, \bibinfo {author} {\bibfnamefont {M.}~\bibnamefont {Borghesi}}, \
  and\ \bibinfo {author} {\bibfnamefont {M.}~\bibnamefont {Passoni}},\ }\href
  {\doibase 10.1103/RevModPhys.85.751} {\bibfield  {journal} {\bibinfo
  {journal} {Reviews of Modern Physics}\ }\textbf {\bibinfo {volume} {85}},\
  \bibinfo {pages} {751} (\bibinfo {year} {2013})}\BibitemShut {NoStop}%
\bibitem [{\citenamefont {Bulanov}\ \emph {et~al.}(2014)\citenamefont
  {Bulanov}, \citenamefont {Wilkens}, \citenamefont {Esirkepov}, \citenamefont
  {Korn}, \citenamefont {Kraft}, \citenamefont {Kraft}, \citenamefont {Molls},\
  and\ \citenamefont {Khoroshkov}}]{Bulanov2014a}%
  \BibitemOpen
  \bibfield  {author} {\bibinfo {author} {\bibfnamefont {S.~V.}\ \bibnamefont
  {Bulanov}}, \bibinfo {author} {\bibfnamefont {J.~J.}\ \bibnamefont
  {Wilkens}}, \bibinfo {author} {\bibfnamefont {T.~Z.}\ \bibnamefont
  {Esirkepov}}, \bibinfo {author} {\bibfnamefont {G.}~\bibnamefont {Korn}},
  \bibinfo {author} {\bibfnamefont {G.}~\bibnamefont {Kraft}}, \bibinfo
  {author} {\bibfnamefont {S.~D.}\ \bibnamefont {Kraft}}, \bibinfo {author}
  {\bibfnamefont {M.}~\bibnamefont {Molls}}, \ and\ \bibinfo {author}
  {\bibfnamefont {V.}~\bibnamefont {Khoroshkov}},\ }\href
  {http://ufn.ru/en/articles/2014/12/a/references.html} {\bibfield  {journal}
  {\bibinfo  {journal} {Physics-Uspekhi}\ }\textbf {\bibinfo {volume} {57}},\
  \bibinfo {pages} {1149} (\bibinfo {year} {2014})}\BibitemShut {NoStop}%
\bibitem [{\citenamefont {Weibel}(1959)}]{Weibel1959}%
  \BibitemOpen
  \bibfield  {author} {\bibinfo {author} {\bibfnamefont {E.}~\bibnamefont
  {Weibel}},\ }\href {\doibase 10.1103/PhysRevLett.2.83} {\bibfield  {journal}
  {\bibinfo  {journal} {Physical Review Letters}\ }\textbf {\bibinfo {volume}
  {2}},\ \bibinfo {pages} {83} (\bibinfo {year} {1959})}\BibitemShut {NoStop}%
\bibitem [{\citenamefont {Fried}(1959)}]{Fried1959}%
  \BibitemOpen
  \bibfield  {author} {\bibinfo {author} {\bibfnamefont {B.~D.}\ \bibnamefont
  {Fried}},\ }\href@noop {} {\bibfield  {journal} {\bibinfo  {journal} {Phys.
  Fluids}\ }\textbf {\bibinfo {volume} {2}},\ \bibinfo {pages} {10} (\bibinfo
  {year} {1959})}\BibitemShut {NoStop}%
\bibitem [{\citenamefont {Bell}(2004)}]{Bell2004}%
  \BibitemOpen
  \bibfield  {author} {\bibinfo {author} {\bibfnamefont {A.~R.}\ \bibnamefont
  {Bell}},\ }\href {\doibase 10.1111/j.1365-2966.2004.08097.x} {\bibfield
  {journal} {\bibinfo  {journal} {Monthly Notices of the Royal Astronomical
  Society}\ }\textbf {\bibinfo {volume} {353}},\ \bibinfo {pages} {550}
  (\bibinfo {year} {2004})}\BibitemShut {NoStop}%
\bibitem [{\citenamefont {Kulsrund}\ and\ \citenamefont
  {Pearce}(1969)}]{Kulsrud1969}%
  \BibitemOpen
  \bibfield  {author} {\bibinfo {author} {\bibfnamefont {R.}~\bibnamefont
  {Kulsrund}}\ and\ \bibinfo {author} {\bibfnamefont {W.~P.}\ \bibnamefont
  {Pearce}},\ }\href@noop {} {\bibfield  {journal} {\bibinfo  {journal} {The
  Astrophysical Journal}\ }\textbf {\bibinfo {volume} {156}},\ \bibinfo {pages}
  {445} (\bibinfo {year} {1969})}\BibitemShut {NoStop}%
\bibitem [{\citenamefont {Wentzel}(1974)}]{Wentzel1974}%
  \BibitemOpen
  \bibfield  {author} {\bibinfo {author} {\bibfnamefont {D.~G.}\ \bibnamefont
  {Wentzel}},\ }\href@noop {} {\bibfield  {journal} {\bibinfo  {journal}
  {ARA{\&}A}\ }\textbf {\bibinfo {volume} {12}},\ \bibinfo {pages} {71}
  (\bibinfo {year} {1974})}\BibitemShut {NoStop}%
\bibitem [{\citenamefont {Macchi}\ \emph {et~al.}(2012)\citenamefont {Macchi},
  \citenamefont {Nindrayog},\ and\ \citenamefont {Pegoraro}}]{Macchi2012}%
  \BibitemOpen
  \bibfield  {author} {\bibinfo {author} {\bibfnamefont {A.}~\bibnamefont
  {Macchi}}, \bibinfo {author} {\bibfnamefont {A.~S.}\ \bibnamefont
  {Nindrayog}}, \ and\ \bibinfo {author} {\bibfnamefont {F.}~\bibnamefont
  {Pegoraro}},\ }\href {\doibase 10.1103/PhysRevE.85.046402} {\bibfield
  {journal} {\bibinfo  {journal} {Physical Review E}\ }\textbf {\bibinfo
  {volume} {85}},\ \bibinfo {pages} {046402} (\bibinfo {year}
  {2012})}\BibitemShut {NoStop}%
\bibitem [{\citenamefont {Haberberger}\ \emph {et~al.}(2012)\citenamefont
  {Haberberger}, \citenamefont {Tochitsky}, \citenamefont {Fiuza},
  \citenamefont {Gong}, \citenamefont {Fonseca}, \citenamefont {Silva},
  \citenamefont {Mori},\ and\ \citenamefont {Joshi}}]{Haberberger2012}%
  \BibitemOpen
  \bibfield  {author} {\bibinfo {author} {\bibfnamefont {D.}~\bibnamefont
  {Haberberger}}, \bibinfo {author} {\bibfnamefont {S.}~\bibnamefont
  {Tochitsky}}, \bibinfo {author} {\bibfnamefont {F.}~\bibnamefont {Fiuza}},
  \bibinfo {author} {\bibfnamefont {C.}~\bibnamefont {Gong}}, \bibinfo {author}
  {\bibfnamefont {R.~A.}\ \bibnamefont {Fonseca}}, \bibinfo {author}
  {\bibfnamefont {L.~O.}\ \bibnamefont {Silva}}, \bibinfo {author}
  {\bibfnamefont {W.~B.}\ \bibnamefont {Mori}}, \ and\ \bibinfo {author}
  {\bibfnamefont {C.}~\bibnamefont {Joshi}},\ }\href {\doibase
  10.1038/nphys2130} {\bibfield  {journal} {\bibinfo  {journal} {Nature
  Physics}\ }\textbf {\bibinfo {volume} {8}},\ \bibinfo {pages} {95} (\bibinfo
  {year} {2012})}\BibitemShut {NoStop}%
\bibitem [{\citenamefont {Fiuza}\ \emph {et~al.}(2012)\citenamefont {Fiuza},
  \citenamefont {Stockem}, \citenamefont {Boella}, \citenamefont {Fonseca},
  \citenamefont {Silva}, \citenamefont {Haberberger}, \citenamefont
  {Tochitsky}, \citenamefont {Gong}, \citenamefont {Mori},\ and\ \citenamefont
  {Joshi}}]{Fiuza2012}%
  \BibitemOpen
  \bibfield  {author} {\bibinfo {author} {\bibfnamefont {F.}~\bibnamefont
  {Fiuza}}, \bibinfo {author} {\bibfnamefont {A.}~\bibnamefont {Stockem}},
  \bibinfo {author} {\bibfnamefont {E.}~\bibnamefont {Boella}}, \bibinfo
  {author} {\bibfnamefont {R.~A.}\ \bibnamefont {Fonseca}}, \bibinfo {author}
  {\bibfnamefont {L.~O.}\ \bibnamefont {Silva}}, \bibinfo {author}
  {\bibfnamefont {D.}~\bibnamefont {Haberberger}}, \bibinfo {author}
  {\bibfnamefont {S.}~\bibnamefont {Tochitsky}}, \bibinfo {author}
  {\bibfnamefont {C.}~\bibnamefont {Gong}}, \bibinfo {author} {\bibfnamefont
  {W.~B.}\ \bibnamefont {Mori}}, \ and\ \bibinfo {author} {\bibfnamefont
  {C.}~\bibnamefont {Joshi}},\ }\href {\doibase 10.1103/PhysRevLett.109.215001}
  {\bibfield  {journal} {\bibinfo  {journal} {Physical Review Letters}\
  }\textbf {\bibinfo {volume} {109}},\ \bibinfo {pages} {215001} (\bibinfo
  {year} {2012})}\BibitemShut {NoStop}%
\bibitem [{\citenamefont {Tresca}\ \emph {et~al.}(2015)\citenamefont {Tresca},
  \citenamefont {Dover}, \citenamefont {Cook}, \citenamefont {Maharjan},
  \citenamefont {Polyanskiy}, \citenamefont {Najmudin}, \citenamefont
  {Shkolnikov},\ and\ \citenamefont {Pogorelsky}}]{Tresca2015}%
  \BibitemOpen
  \bibfield  {author} {\bibinfo {author} {\bibfnamefont {O.}~\bibnamefont
  {Tresca}}, \bibinfo {author} {\bibfnamefont {N.~P.}\ \bibnamefont {Dover}},
  \bibinfo {author} {\bibfnamefont {N.}~\bibnamefont {Cook}}, \bibinfo {author}
  {\bibfnamefont {C.}~\bibnamefont {Maharjan}}, \bibinfo {author}
  {\bibfnamefont {M.~N.}\ \bibnamefont {Polyanskiy}}, \bibinfo {author}
  {\bibfnamefont {Z.}~\bibnamefont {Najmudin}}, \bibinfo {author}
  {\bibfnamefont {P.}~\bibnamefont {Shkolnikov}}, \ and\ \bibinfo {author}
  {\bibfnamefont {I.}~\bibnamefont {Pogorelsky}},\ }\href {\doibase
  10.1103/PhysRevLett.115.094802} {\bibfield  {journal} {\bibinfo  {journal}
  {Physical Review Letters}\ }\textbf {\bibinfo {volume} {115}},\ \bibinfo
  {pages} {094802} (\bibinfo {year} {2015})}\BibitemShut {NoStop}%
\bibitem [{\citenamefont {Antici}\ \emph {et~al.}(2017)\citenamefont {Antici},
  \citenamefont {Boella}, \citenamefont {Chen}, \citenamefont {Andrews},
  \citenamefont {Barberio}, \citenamefont {B{\"{o}}ker}, \citenamefont
  {Cardelli}, \citenamefont {Feugeas}, \citenamefont {Glesser}, \citenamefont
  {Nicola{\"{i}}}, \citenamefont {Romagnani}, \citenamefont {Scisci{\`{o}}},
  \citenamefont {Starodubtsev}, \citenamefont {Willi}, \citenamefont {Kieffer},
  \citenamefont {Tikhonchuk}, \citenamefont {P{\'{e}}pin}, \citenamefont
  {Silva}, \citenamefont {Humi{\`{e}}res},\ and\ \citenamefont
  {Fuchs}}]{Antici2017}%
  \BibitemOpen
  \bibfield  {author} {\bibinfo {author} {\bibfnamefont {P.}~\bibnamefont
  {Antici}}, \bibinfo {author} {\bibfnamefont {E.}~\bibnamefont {Boella}},
  \bibinfo {author} {\bibfnamefont {S.~N.}\ \bibnamefont {Chen}}, \bibinfo
  {author} {\bibfnamefont {D.~S.}\ \bibnamefont {Andrews}}, \bibinfo {author}
  {\bibfnamefont {M.}~\bibnamefont {Barberio}}, \bibinfo {author}
  {\bibfnamefont {J.}~\bibnamefont {B{\"{o}}ker}}, \bibinfo {author}
  {\bibfnamefont {F.}~\bibnamefont {Cardelli}}, \bibinfo {author}
  {\bibfnamefont {J.~L.}\ \bibnamefont {Feugeas}}, \bibinfo {author}
  {\bibfnamefont {M.}~\bibnamefont {Glesser}}, \bibinfo {author} {\bibfnamefont
  {P.}~\bibnamefont {Nicola{\"{i}}}}, \bibinfo {author} {\bibfnamefont
  {L.}~\bibnamefont {Romagnani}}, \bibinfo {author} {\bibfnamefont
  {M.}~\bibnamefont {Scisci{\`{o}}}}, \bibinfo {author} {\bibfnamefont
  {M.}~\bibnamefont {Starodubtsev}}, \bibinfo {author} {\bibfnamefont
  {O.}~\bibnamefont {Willi}}, \bibinfo {author} {\bibfnamefont {J.~C.}\
  \bibnamefont {Kieffer}}, \bibinfo {author} {\bibfnamefont {V.}~\bibnamefont
  {Tikhonchuk}}, \bibinfo {author} {\bibfnamefont {H.}~\bibnamefont
  {P{\'{e}}pin}}, \bibinfo {author} {\bibfnamefont {L.~O.}\ \bibnamefont
  {Silva}}, \bibinfo {author} {\bibfnamefont {E.~D.}\ \bibnamefont
  {Humi{\`{e}}res}}, \ and\ \bibinfo {author} {\bibfnamefont {J.}~\bibnamefont
  {Fuchs}},\ }\href {\doibase 10.1038/s41598-017-15449-8} {\bibfield  {journal}
  {\bibinfo  {journal} {Scientific Reports}\ }\textbf {\bibinfo {volume} {7}},\
  \bibinfo {pages} {16463} (\bibinfo {year} {2017})},\ \Eprint
  {http://arxiv.org/abs/1708.02539} {arXiv:1708.02539} \BibitemShut {NoStop}%
\bibitem [{\citenamefont {Zhang}\ \emph {et~al.}(2017)\citenamefont {Zhang},
  \citenamefont {Shen}, \citenamefont {Wang}, \citenamefont {Zhai},
  \citenamefont {Li}, \citenamefont {Lu}, \citenamefont {Li}, \citenamefont
  {Xu}, \citenamefont {Wang}, \citenamefont {Liang}, \citenamefont {Leng},
  \citenamefont {Li},\ and\ \citenamefont {Xu}}]{Zhang2017}%
  \BibitemOpen
  \bibfield  {author} {\bibinfo {author} {\bibfnamefont {H.}~\bibnamefont
  {Zhang}}, \bibinfo {author} {\bibfnamefont {B.~F.}\ \bibnamefont {Shen}},
  \bibinfo {author} {\bibfnamefont {W.~P.}\ \bibnamefont {Wang}}, \bibinfo
  {author} {\bibfnamefont {S.~H.}\ \bibnamefont {Zhai}}, \bibinfo {author}
  {\bibfnamefont {S.~S.}\ \bibnamefont {Li}}, \bibinfo {author} {\bibfnamefont
  {X.~M.}\ \bibnamefont {Lu}}, \bibinfo {author} {\bibfnamefont {J.~F.}\
  \bibnamefont {Li}}, \bibinfo {author} {\bibfnamefont {R.~J.}\ \bibnamefont
  {Xu}}, \bibinfo {author} {\bibfnamefont {X.~L.}\ \bibnamefont {Wang}},
  \bibinfo {author} {\bibfnamefont {X.~Y.}\ \bibnamefont {Liang}}, \bibinfo
  {author} {\bibfnamefont {Y.~X.}\ \bibnamefont {Leng}}, \bibinfo {author}
  {\bibfnamefont {R.~X.}\ \bibnamefont {Li}}, \ and\ \bibinfo {author}
  {\bibfnamefont {Z.~Z.}\ \bibnamefont {Xu}},\ }\href {\doibase
  10.1103/PhysRevLett.119.164801} {\bibfield  {journal} {\bibinfo  {journal}
  {Physical Review Letters}\ }\textbf {\bibinfo {volume} {119}},\ \bibinfo
  {pages} {164801} (\bibinfo {year} {2017})}\BibitemShut {NoStop}%
\bibitem [{\citenamefont {Chen}\ \emph {et~al.}(2017)\citenamefont {Chen},
  \citenamefont {Vranic}, \citenamefont {Gangolf}, \citenamefont {Boella},
  \citenamefont {Antici}, \citenamefont {Bailly-Grandvaux}, \citenamefont
  {Loiseau}, \citenamefont {P{\'{e}}pin}, \citenamefont {Revet}, \citenamefont
  {Santos}, \citenamefont {Schroer}, \citenamefont {Starodubtsev},
  \citenamefont {Willi}, \citenamefont {Silva}, \citenamefont
  {D'humi{\`{e}}res},\ and\ \citenamefont {Fuchs}}]{Chen2017a}%
  \BibitemOpen
  \bibfield  {author} {\bibinfo {author} {\bibfnamefont {S.~N.}\ \bibnamefont
  {Chen}}, \bibinfo {author} {\bibfnamefont {M.}~\bibnamefont {Vranic}},
  \bibinfo {author} {\bibfnamefont {T.}~\bibnamefont {Gangolf}}, \bibinfo
  {author} {\bibfnamefont {E.}~\bibnamefont {Boella}}, \bibinfo {author}
  {\bibfnamefont {P.}~\bibnamefont {Antici}}, \bibinfo {author} {\bibfnamefont
  {M.}~\bibnamefont {Bailly-Grandvaux}}, \bibinfo {author} {\bibfnamefont
  {P.}~\bibnamefont {Loiseau}}, \bibinfo {author} {\bibfnamefont
  {H.}~\bibnamefont {P{\'{e}}pin}}, \bibinfo {author} {\bibfnamefont
  {G.}~\bibnamefont {Revet}}, \bibinfo {author} {\bibfnamefont {J.~J.}\
  \bibnamefont {Santos}}, \bibinfo {author} {\bibfnamefont {A.~M.}\
  \bibnamefont {Schroer}}, \bibinfo {author} {\bibfnamefont {M.}~\bibnamefont
  {Starodubtsev}}, \bibinfo {author} {\bibfnamefont {O.}~\bibnamefont {Willi}},
  \bibinfo {author} {\bibfnamefont {L.~O.}\ \bibnamefont {Silva}}, \bibinfo
  {author} {\bibfnamefont {E.}~\bibnamefont {D'humi{\`{e}}res}}, \ and\
  \bibinfo {author} {\bibfnamefont {J.}~\bibnamefont {Fuchs}},\ }\href
  {\doibase 10.1038/s41598-017-12910-6} {\bibfield  {journal} {\bibinfo
  {journal} {Scientific Reports}\ }\textbf {\bibinfo {volume} {7}},\ \bibinfo
  {pages} {13505} (\bibinfo {year} {2017})}\BibitemShut {NoStop}%
\bibitem [{\citenamefont {Pak}\ \emph {et~al.}(2018)\citenamefont {Pak},
  \citenamefont {Kerr}, \citenamefont {Lemos}, \citenamefont {Link},
  \citenamefont {Patel}, \citenamefont {Albert}, \citenamefont {Divol},
  \citenamefont {Pollock}, \citenamefont {Haberberger}, \citenamefont {Froula},
  \citenamefont {Gauthier}, \citenamefont {Glenzer}, \citenamefont {Longman},
  \citenamefont {Manzoor}, \citenamefont {Fedosejevs}, \citenamefont
  {Tochitsky}, \citenamefont {Joshi},\ and\ \citenamefont {Fiuza}}]{Pak2018a}%
  \BibitemOpen
  \bibfield  {author} {\bibinfo {author} {\bibfnamefont {A.}~\bibnamefont
  {Pak}}, \bibinfo {author} {\bibfnamefont {S.}~\bibnamefont {Kerr}}, \bibinfo
  {author} {\bibfnamefont {N.}~\bibnamefont {Lemos}}, \bibinfo {author}
  {\bibfnamefont {A.}~\bibnamefont {Link}}, \bibinfo {author} {\bibfnamefont
  {P.}~\bibnamefont {Patel}}, \bibinfo {author} {\bibfnamefont
  {F.}~\bibnamefont {Albert}}, \bibinfo {author} {\bibfnamefont
  {L.}~\bibnamefont {Divol}}, \bibinfo {author} {\bibfnamefont {B.~B.}\
  \bibnamefont {Pollock}}, \bibinfo {author} {\bibfnamefont {D.}~\bibnamefont
  {Haberberger}}, \bibinfo {author} {\bibfnamefont {D.}~\bibnamefont {Froula}},
  \bibinfo {author} {\bibfnamefont {M.}~\bibnamefont {Gauthier}}, \bibinfo
  {author} {\bibfnamefont {S.~H.}\ \bibnamefont {Glenzer}}, \bibinfo {author}
  {\bibfnamefont {A.}~\bibnamefont {Longman}}, \bibinfo {author} {\bibfnamefont
  {L.}~\bibnamefont {Manzoor}}, \bibinfo {author} {\bibfnamefont
  {R.}~\bibnamefont {Fedosejevs}}, \bibinfo {author} {\bibfnamefont
  {S.}~\bibnamefont {Tochitsky}}, \bibinfo {author} {\bibfnamefont
  {C.}~\bibnamefont {Joshi}}, \ and\ \bibinfo {author} {\bibfnamefont
  {F.}~\bibnamefont {Fiuza}},\ }\href {\doibase
  10.1103/PhysRevAccelBeams.21.103401} {\bibfield  {journal} {\bibinfo
  {journal} {Physical Review Accelerators and Beams}\ }\textbf {\bibinfo
  {volume} {21}},\ \bibinfo {pages} {103401} (\bibinfo {year} {2018})},\
  \Eprint {http://arxiv.org/abs/1810.08190} {arXiv:1810.08190} \BibitemShut
  {NoStop}%
\bibitem [{\citenamefont {Matsui}\ \emph {et~al.}(2019)\citenamefont {Matsui},
  \citenamefont {Fukuda},\ and\ \citenamefont {Kishimoto}}]{Matsui2019}%
  \BibitemOpen
  \bibfield  {author} {\bibinfo {author} {\bibfnamefont {R.}~\bibnamefont
  {Matsui}}, \bibinfo {author} {\bibfnamefont {Y.}~\bibnamefont {Fukuda}}, \
  and\ \bibinfo {author} {\bibfnamefont {Y.}~\bibnamefont {Kishimoto}},\ }\href
  {\doibase 10.1103/PhysRevLett.122.014804} {\bibfield  {journal} {\bibinfo
  {journal} {Physical Review Letters}\ }\textbf {\bibinfo {volume} {122}},\
  \bibinfo {pages} {014804} (\bibinfo {year} {2019})}\BibitemShut {NoStop}%
\bibitem [{\citenamefont {Ota}\ \emph {et~al.}(2019)\citenamefont {Ota},
  \citenamefont {Morace}, \citenamefont {Kumar}, \citenamefont {Kambayashi},
  \citenamefont {Egashira}, \citenamefont {Kanasaki}, \citenamefont {Fukuda},\
  and\ \citenamefont {Sakawa}}]{Ota2019}%
  \BibitemOpen
  \bibfield  {author} {\bibinfo {author} {\bibfnamefont {M.}~\bibnamefont
  {Ota}}, \bibinfo {author} {\bibfnamefont {A.}~\bibnamefont {Morace}},
  \bibinfo {author} {\bibfnamefont {R.}~\bibnamefont {Kumar}}, \bibinfo
  {author} {\bibfnamefont {S.}~\bibnamefont {Kambayashi}}, \bibinfo {author}
  {\bibfnamefont {S.}~\bibnamefont {Egashira}}, \bibinfo {author}
  {\bibfnamefont {M.}~\bibnamefont {Kanasaki}}, \bibinfo {author}
  {\bibfnamefont {Y.}~\bibnamefont {Fukuda}}, \ and\ \bibinfo {author}
  {\bibfnamefont {Y.}~\bibnamefont {Sakawa}},\ }\href {\doibase
  10.1016/j.hedp.2019.100697} {\bibfield  {journal} {\bibinfo  {journal} {High
  Energy Density Physics}\ }\textbf {\bibinfo {volume} {33}},\ \bibinfo {pages}
  {100697} (\bibinfo {year} {2019})}\BibitemShut {NoStop}%
\bibitem [{\citenamefont {Singh}\ \emph {et~al.}(2020)\citenamefont {Singh},
  \citenamefont {Pathak}, \citenamefont {Shin}, \citenamefont {Choi},
  \citenamefont {Nakajima}, \citenamefont {Lee}, \citenamefont {Sung},
  \citenamefont {Lee}, \citenamefont {Rhee}, \citenamefont {Aniculaesei},
  \citenamefont {Kim}, \citenamefont {Pae}, \citenamefont {Cho}, \citenamefont
  {Hojbota}, \citenamefont {Lee}, \citenamefont {Mollica}, \citenamefont
  {Malka}, \citenamefont {Ryu}, \citenamefont {Kim},\ and\ \citenamefont
  {Nam}}]{Singh2020}%
  \BibitemOpen
  \bibfield  {author} {\bibinfo {author} {\bibfnamefont {P.~K.}\ \bibnamefont
  {Singh}}, \bibinfo {author} {\bibfnamefont {V.~B.}\ \bibnamefont {Pathak}},
  \bibinfo {author} {\bibfnamefont {J.~H.}\ \bibnamefont {Shin}}, \bibinfo
  {author} {\bibfnamefont {I.~W.}\ \bibnamefont {Choi}}, \bibinfo {author}
  {\bibfnamefont {K.}~\bibnamefont {Nakajima}}, \bibinfo {author}
  {\bibfnamefont {S.~K.}\ \bibnamefont {Lee}}, \bibinfo {author} {\bibfnamefont
  {J.~H.}\ \bibnamefont {Sung}}, \bibinfo {author} {\bibfnamefont {H.~W.}\
  \bibnamefont {Lee}}, \bibinfo {author} {\bibfnamefont {Y.~J.}\ \bibnamefont
  {Rhee}}, \bibinfo {author} {\bibfnamefont {C.}~\bibnamefont {Aniculaesei}},
  \bibinfo {author} {\bibfnamefont {C.~M.}\ \bibnamefont {Kim}}, \bibinfo
  {author} {\bibfnamefont {K.~H.}\ \bibnamefont {Pae}}, \bibinfo {author}
  {\bibfnamefont {M.~H.}\ \bibnamefont {Cho}}, \bibinfo {author} {\bibfnamefont
  {C.}~\bibnamefont {Hojbota}}, \bibinfo {author} {\bibfnamefont {S.~G.}\
  \bibnamefont {Lee}}, \bibinfo {author} {\bibfnamefont {F.}~\bibnamefont
  {Mollica}}, \bibinfo {author} {\bibfnamefont {V.}~\bibnamefont {Malka}},
  \bibinfo {author} {\bibfnamefont {C.~M.}\ \bibnamefont {Ryu}}, \bibinfo
  {author} {\bibfnamefont {H.~T.}\ \bibnamefont {Kim}}, \ and\ \bibinfo
  {author} {\bibfnamefont {C.~H.}\ \bibnamefont {Nam}},\ }\href {\doibase
  10.1038/s41598-020-75455-1} {\bibfield  {journal} {\bibinfo  {journal}
  {Scientific Reports}\ }\textbf {\bibinfo {volume} {10}},\ \bibinfo {pages}
  {18452} (\bibinfo {year} {2020})}\BibitemShut {NoStop}%
\bibitem [{\citenamefont {Tochitsky}\ \emph {et~al.}(2020)\citenamefont
  {Tochitsky}, \citenamefont {Pak}, \citenamefont {Fiuza}, \citenamefont
  {Haberberger}, \citenamefont {Lemos}, \citenamefont {Link}, \citenamefont
  {Froula},\ and\ \citenamefont {Joshi}}]{Tochitsky2020}%
  \BibitemOpen
  \bibfield  {author} {\bibinfo {author} {\bibfnamefont {S.}~\bibnamefont
  {Tochitsky}}, \bibinfo {author} {\bibfnamefont {A.}~\bibnamefont {Pak}},
  \bibinfo {author} {\bibfnamefont {F.}~\bibnamefont {Fiuza}}, \bibinfo
  {author} {\bibfnamefont {D.}~\bibnamefont {Haberberger}}, \bibinfo {author}
  {\bibfnamefont {N.}~\bibnamefont {Lemos}}, \bibinfo {author} {\bibfnamefont
  {A.}~\bibnamefont {Link}}, \bibinfo {author} {\bibfnamefont {D.~H.}\
  \bibnamefont {Froula}}, \ and\ \bibinfo {author} {\bibfnamefont
  {C.}~\bibnamefont {Joshi}},\ }\href {\doibase 10.1063/1.5144446} {\bibfield
  {journal} {\bibinfo  {journal} {Physics of Plasmas}\ }\textbf {\bibinfo
  {volume} {27}},\ \bibinfo {pages} {083102} (\bibinfo {year}
  {2020})}\BibitemShut {NoStop}%
\bibitem [{\citenamefont {Boella}\ \emph {et~al.}(2021)\citenamefont {Boella},
  \citenamefont {Bingham}, \citenamefont {Cairns}, \citenamefont {Norreys},
  \citenamefont {Trines}, \citenamefont {Scott}, \citenamefont {Vranic},
  \citenamefont {Shukla},\ and\ \citenamefont {Silva}}]{Boella2021}%
  \BibitemOpen
  \bibfield  {author} {\bibinfo {author} {\bibfnamefont {E.}~\bibnamefont
  {Boella}}, \bibinfo {author} {\bibfnamefont {R.}~\bibnamefont {Bingham}},
  \bibinfo {author} {\bibfnamefont {R.~A.}\ \bibnamefont {Cairns}}, \bibinfo
  {author} {\bibfnamefont {P.}~\bibnamefont {Norreys}}, \bibinfo {author}
  {\bibfnamefont {R.}~\bibnamefont {Trines}}, \bibinfo {author} {\bibfnamefont
  {R.}~\bibnamefont {Scott}}, \bibinfo {author} {\bibfnamefont
  {M.}~\bibnamefont {Vranic}}, \bibinfo {author} {\bibfnamefont
  {N.}~\bibnamefont {Shukla}}, \ and\ \bibinfo {author} {\bibfnamefont {L.~O.}\
  \bibnamefont {Silva}},\ }\href {\doibase 10.1098/rsta.2020.0039rsta20200039}
  {\bibfield  {journal} {\bibinfo  {journal} {Philosophical Transactions of the
  Royal Society A: Mathematical, Physical and Engineering Sciences}\ }\textbf
  {\bibinfo {volume} {379}},\ \bibinfo {pages} {20200039} (\bibinfo {year}
  {2021})}\BibitemShut {NoStop}%
\bibitem [{\citenamefont {Kumar}\ \emph {et~al.}(2019)\citenamefont {Kumar},
  \citenamefont {Sakawa}, \citenamefont {D{\"{o}}hl}, \citenamefont {Woolsey},\
  and\ \citenamefont {Morace}}]{Kumar2019a}%
  \BibitemOpen
  \bibfield  {author} {\bibinfo {author} {\bibfnamefont {R.}~\bibnamefont
  {Kumar}}, \bibinfo {author} {\bibfnamefont {Y.}~\bibnamefont {Sakawa}},
  \bibinfo {author} {\bibfnamefont {L.~N.~K.}\ \bibnamefont {D{\"{o}}hl}},
  \bibinfo {author} {\bibfnamefont {N.}~\bibnamefont {Woolsey}}, \ and\
  \bibinfo {author} {\bibfnamefont {A.}~\bibnamefont {Morace}},\ }\href
  {\doibase 10.1103/PhysRevAccelBeams.22.043401} {\bibfield  {journal}
  {\bibinfo  {journal} {Physical Review Accelerators and Beams}\ }\textbf
  {\bibinfo {volume} {22}},\ \bibinfo {pages} {043401} (\bibinfo {year}
  {2019})}\BibitemShut {NoStop}%
\bibitem [{\citenamefont {Kumar}\ \emph {et~al.}(2021)\citenamefont {Kumar},
  \citenamefont {Sakawa}, \citenamefont {Sano}, \citenamefont {D{\"{o}}hl},
  \citenamefont {Woolsey},\ and\ \citenamefont {Morace}}]{Kumar2021}%
  \BibitemOpen
  \bibfield  {author} {\bibinfo {author} {\bibfnamefont {R.}~\bibnamefont
  {Kumar}}, \bibinfo {author} {\bibfnamefont {Y.}~\bibnamefont {Sakawa}},
  \bibinfo {author} {\bibfnamefont {T.}~\bibnamefont {Sano}}, \bibinfo {author}
  {\bibfnamefont {L.~N.~K.}\ \bibnamefont {D{\"{o}}hl}}, \bibinfo {author}
  {\bibfnamefont {N.}~\bibnamefont {Woolsey}}, \ and\ \bibinfo {author}
  {\bibfnamefont {A.}~\bibnamefont {Morace}},\ }\href {\doibase
  10.1103/PhysRevE.103.043201} {\bibfield  {journal} {\bibinfo  {journal}
  {Physical Review E}\ }\textbf {\bibinfo {volume} {103}},\ \bibinfo {pages}
  {043201} (\bibinfo {year} {2021})},\ \Eprint
  {http://arxiv.org/abs/2104.00866} {arXiv:2104.00866} \BibitemShut {NoStop}%
\bibitem [{\citenamefont {Sakawa}\ \emph {et~al.}(2021)\citenamefont {Sakawa},
  \citenamefont {Ohira}, \citenamefont {Kumar}, \citenamefont {Morace},
  \citenamefont {D{\"{o}}hl},\ and\ \citenamefont {Woolsey}}]{Sakawa2021}%
  \BibitemOpen
  \bibfield  {author} {\bibinfo {author} {\bibfnamefont {Y.}~\bibnamefont
  {Sakawa}}, \bibinfo {author} {\bibfnamefont {Y.}~\bibnamefont {Ohira}},
  \bibinfo {author} {\bibfnamefont {R.}~\bibnamefont {Kumar}}, \bibinfo
  {author} {\bibfnamefont {A.}~\bibnamefont {Morace}}, \bibinfo {author}
  {\bibfnamefont {L.~N.~K.}\ \bibnamefont {D{\"{o}}hl}}, \ and\ \bibinfo
  {author} {\bibfnamefont {N.}~\bibnamefont {Woolsey}},\ }\href {\doibase
  10.1103/physreve.104.055202} {\bibfield  {journal} {\bibinfo  {journal}
  {Physical Review E}\ }\textbf {\bibinfo {volume} {104}},\ \bibinfo {pages}
  {055202} (\bibinfo {year} {2021})},\ \Eprint
  {http://arxiv.org/abs/2110.07175} {arXiv:2110.07175} \BibitemShut {NoStop}%
\bibitem [{\citenamefont {Mora}(2003)}]{Mora2003}%
  \BibitemOpen
  \bibfield  {author} {\bibinfo {author} {\bibfnamefont {P.}~\bibnamefont
  {Mora}},\ }\href {\doibase 10.1103/PhysRevLett.90.185002} {\bibfield
  {journal} {\bibinfo  {journal} {Physical Review Letters}\ }\textbf {\bibinfo
  {volume} {90}},\ \bibinfo {pages} {185002} (\bibinfo {year}
  {2003})}\BibitemShut {NoStop}%
  %
  %
  \bibitem{supple}
See Supplemental Material for additional data.
%
 %
\bibitem [{\citenamefont {Faenov}\ \emph {et~al.}(1994)\citenamefont {Faenov},
  \citenamefont {Pikuz}, \citenamefont {Erko}, \citenamefont {Bryunetkin},
  \citenamefont {Dyakin}, \citenamefont {Ivanenkov}, \citenamefont {Mingaleev},
  \citenamefont {Pikuz}, \citenamefont {Romanova},\ and\ \citenamefont
  {Shelkovenko}}]{Faenov1994}%
  \BibitemOpen
  \bibfield  {author} {\bibinfo {author} {\bibfnamefont {A.~Y.}\ \bibnamefont
  {Faenov}}, \bibinfo {author} {\bibfnamefont {S.~A.}\ \bibnamefont {Pikuz}},
  \bibinfo {author} {\bibfnamefont {A.~I.}\ \bibnamefont {Erko}}, \bibinfo
  {author} {\bibfnamefont {B.~A.}\ \bibnamefont {Bryunetkin}}, \bibinfo
  {author} {\bibfnamefont {V.~M.}\ \bibnamefont {Dyakin}}, \bibinfo {author}
  {\bibfnamefont {G.~V.}\ \bibnamefont {Ivanenkov}}, \bibinfo {author}
  {\bibfnamefont {A.~R.}\ \bibnamefont {Mingaleev}}, \bibinfo {author}
  {\bibfnamefont {T.~A.}\ \bibnamefont {Pikuz}}, \bibinfo {author}
  {\bibfnamefont {V.~M.}\ \bibnamefont {Romanova}}, \ and\ \bibinfo {author}
  {\bibfnamefont {T.~A.}\ \bibnamefont {Shelkovenko}},\ }\href {\doibase
  10.1088/0031-8949/50/4/003} {\bibfield  {journal} {\bibinfo  {journal}
  {Physica Scripta}\ }\textbf {\bibinfo {volume} {50}},\ \bibinfo {pages} {333}
  (\bibinfo {year} {1994})}\BibitemShut {NoStop}%
\bibitem [{\citenamefont {Kuramitsu}\ \emph {et~al.}(2022)\citenamefont
  {Kuramitsu}, \citenamefont {Minami}, \citenamefont {Hihara}, \citenamefont
  {Sakai}, \citenamefont {Nishimoto}, \citenamefont {Isayama}, \citenamefont
  {Liao}, \citenamefont {Wu}, \citenamefont {Woon}, \citenamefont {Chen},
  \citenamefont {Liu}, \citenamefont {He}, \citenamefont {Su}, \citenamefont
  {Ota}, \citenamefont {Egashira}, \citenamefont {Morace}, \citenamefont
  {Sakawa}, \citenamefont {Abe}, \citenamefont {Habara}, \citenamefont
  {Kodama}, \citenamefont {D{\"{o}}hl}, \citenamefont {Woolsey}, \citenamefont
  {Koenig}, \citenamefont {Kumar}, \citenamefont {Ohnishi}, \citenamefont
  {Kanasaki}, \citenamefont {Asai}, \citenamefont {Yamauchi}, \citenamefont
  {Oda}, \citenamefont {Kondo}, \citenamefont {Kiriyama},\ and\ \citenamefont
  {Fukuda}}]{Kuramitsu2022}%
  \BibitemOpen
  \bibfield  {author} {\bibinfo {author} {\bibfnamefont {Y.}~\bibnamefont
  {Kuramitsu}}, \bibinfo {author} {\bibfnamefont {T.}~\bibnamefont {Minami}},
  \bibinfo {author} {\bibfnamefont {T.}~\bibnamefont {Hihara}}, \bibinfo
  {author} {\bibfnamefont {K.}~\bibnamefont {Sakai}}, \bibinfo {author}
  {\bibfnamefont {T.}~\bibnamefont {Nishimoto}}, \bibinfo {author}
  {\bibfnamefont {S.}~\bibnamefont {Isayama}}, \bibinfo {author} {\bibfnamefont
  {Y.~T.}\ \bibnamefont {Liao}}, \bibinfo {author} {\bibfnamefont {K.~T.}\
  \bibnamefont {Wu}}, \bibinfo {author} {\bibfnamefont {W.~Y.}\ \bibnamefont
  {Woon}}, \bibinfo {author} {\bibfnamefont {S.~H.}\ \bibnamefont {Chen}},
  \bibinfo {author} {\bibfnamefont {Y.~L.}\ \bibnamefont {Liu}}, \bibinfo
  {author} {\bibfnamefont {S.~M.}\ \bibnamefont {He}}, \bibinfo {author}
  {\bibfnamefont {C.~Y.}\ \bibnamefont {Su}}, \bibinfo {author} {\bibfnamefont
  {M.}~\bibnamefont {Ota}}, \bibinfo {author} {\bibfnamefont {S.}~\bibnamefont
  {Egashira}}, \bibinfo {author} {\bibfnamefont {A.}~\bibnamefont {Morace}},
  \bibinfo {author} {\bibfnamefont {Y.}~\bibnamefont {Sakawa}}, \bibinfo
  {author} {\bibfnamefont {Y.}~\bibnamefont {Abe}}, \bibinfo {author}
  {\bibfnamefont {H.}~\bibnamefont {Habara}}, \bibinfo {author} {\bibfnamefont
  {R.}~\bibnamefont {Kodama}}, \bibinfo {author} {\bibfnamefont {L.~N.}\
  \bibnamefont {D{\"{o}}hl}}, \bibinfo {author} {\bibfnamefont
  {N.}~\bibnamefont {Woolsey}}, \bibinfo {author} {\bibfnamefont
  {M.}~\bibnamefont {Koenig}}, \bibinfo {author} {\bibfnamefont {H.~S.}\
  \bibnamefont {Kumar}}, \bibinfo {author} {\bibfnamefont {N.}~\bibnamefont
  {Ohnishi}}, \bibinfo {author} {\bibfnamefont {M.}~\bibnamefont {Kanasaki}},
  \bibinfo {author} {\bibfnamefont {T.}~\bibnamefont {Asai}}, \bibinfo {author}
  {\bibfnamefont {T.}~\bibnamefont {Yamauchi}}, \bibinfo {author}
  {\bibfnamefont {K.}~\bibnamefont {Oda}}, \bibinfo {author} {\bibfnamefont
  {K.}~\bibnamefont {Kondo}}, \bibinfo {author} {\bibfnamefont
  {H.}~\bibnamefont {Kiriyama}}, \ and\ \bibinfo {author} {\bibfnamefont
  {Y.}~\bibnamefont {Fukuda}},\ }\href {\doibase 10.1038/s41598-022-06055-4}
  {\bibfield  {journal} {\bibinfo  {journal} {Scientific reports}\ }\textbf
  {\bibinfo {volume} {12}},\ \bibinfo {pages} {2346} (\bibinfo {year}
  {2022})}\BibitemShut {NoStop}%
\bibitem [{Pri()}]{PrismSPECT}%
  \BibitemOpen
  \href {\doibase https://www.prism-cs.com/Software/PrismSPECT.html} {\
  https://www.prism-cs.com/Software/PrismSPECT.html}\BibitemShut {NoStop}%
\bibitem [{\citenamefont {MacFarlane}\ \emph {et~al.}(2004)\citenamefont
  {MacFarlane}, \citenamefont {Golovkin}, \citenamefont {Woodruff},
  \citenamefont {Welch}, \citenamefont {Oliver}, \citenamefont {Mehlhorn},\
  and\ \citenamefont {Campbell}}]{MacFarlane2004}%
  \BibitemOpen
  \bibfield  {author} {\bibinfo {author} {\bibfnamefont {J.~J.}\ \bibnamefont
  {MacFarlane}}, \bibinfo {author} {\bibfnamefont {I.~E.}\ \bibnamefont
  {Golovkin}}, \bibinfo {author} {\bibfnamefont {P.~R.}\ \bibnamefont
  {Woodruff}}, \bibinfo {author} {\bibfnamefont {D.~R.}\ \bibnamefont {Welch}},
  \bibinfo {author} {\bibfnamefont {B.~V.}\ \bibnamefont {Oliver}}, \bibinfo
  {author} {\bibfnamefont {T.~A.}\ \bibnamefont {Mehlhorn}}, \ and\ \bibinfo
  {author} {\bibfnamefont {R.~B.}\ \bibnamefont {Campbell}},\ }\href@noop {}
  {\bibfield  {journal} {\bibinfo  {journal} {Inertial Fusion Sciences and
  Applications 2003: State of the Art, B. A. Hammel, D. D. Meyerhofer, J.
  Meyer- terVehn, and H. Azechi, eds. (American Nuclear Society, La Grange Park, IL)}\ ,\ \bibinfo
  {pages} {457}} (\bibinfo {year} {2004})}\BibitemShut {NoStop}%
\bibitem [{\citenamefont {Arber}\ \emph {et~al.}(2015)\citenamefont {Arber},
  \citenamefont {Bennett}, \citenamefont {Brady}, \citenamefont
  {Lawrence-Douglas}, \citenamefont {Ramsay}, \citenamefont {Sircombe},
  \citenamefont {Gillies}, \citenamefont {Evans}, \citenamefont {Schmitz},
  \citenamefont {Bell},\ and\ \citenamefont {Ridgers}}]{Arber2015}%
  \BibitemOpen
  \bibfield  {author} {\bibinfo {author} {\bibfnamefont {T.~D.}\ \bibnamefont
  {Arber}}, \bibinfo {author} {\bibfnamefont {K.}~\bibnamefont {Bennett}},
  \bibinfo {author} {\bibfnamefont {C.~S.}\ \bibnamefont {Brady}}, \bibinfo
  {author} {\bibfnamefont {A.}~\bibnamefont {Lawrence-Douglas}}, \bibinfo
  {author} {\bibfnamefont {M.~G.}\ \bibnamefont {Ramsay}}, \bibinfo {author}
  {\bibfnamefont {N.~J.}\ \bibnamefont {Sircombe}}, \bibinfo {author}
  {\bibfnamefont {P.}~\bibnamefont {Gillies}}, \bibinfo {author} {\bibfnamefont
  {R.~G.}\ \bibnamefont {Evans}}, \bibinfo {author} {\bibfnamefont
  {H.}~\bibnamefont {Schmitz}}, \bibinfo {author} {\bibfnamefont {A.~R.}\
  \bibnamefont {Bell}}, \ and\ \bibinfo {author} {\bibfnamefont {C.~P.}\
  \bibnamefont {Ridgers}},\ }\href {\doibase 10.1088/0741-3335/57/11/113001}
  {\bibfield  {journal} {\bibinfo  {journal} {Plasma Physics and Controlled
  Fusion}\ }\textbf {\bibinfo {volume} {57}} (\bibinfo {year} {2015}),\
  10.1088/0741-3335/57/11/113001}\BibitemShut {NoStop}%
\bibitem [{\citenamefont {Tidman}\ and\ \citenamefont {Krall}(1971)}]{Tidman}%
  \BibitemOpen
  \bibfield  {author} {\bibinfo {author} {\bibfnamefont {D.~A.}\ \bibnamefont
  {Tidman}}\ and\ \bibinfo {author} {\bibfnamefont {N.~A.}\ \bibnamefont
  {Krall}},\ }\href@noop {} {\emph {\bibinfo {title} {{Shock Waves in
  Collisionless Plasmas}}}}\ (\bibinfo  {publisher} {Wiley-Interscience},\
  \bibinfo {address} {New York},\ \bibinfo {year} {1971})\BibitemShut {NoStop}%
\bibitem [{\citenamefont {Fuchs}\ \emph {et~al.}(2007)\citenamefont {Fuchs},
  \citenamefont {Cecchetti}, \citenamefont {Borghesi}, \citenamefont
  {Grismayer}, \citenamefont {d'Humi{\`{e}}res}, \citenamefont {Antici},
  \citenamefont {Atzeni}, \citenamefont {Mora}, \citenamefont {Pipahl},
  \citenamefont {Romagnani}, \citenamefont {Schiavi}, \citenamefont {Sentoku},
  \citenamefont {Toncian}, \citenamefont {Audebert},\ and\ \citenamefont
  {Willi}}]{Fuchs2007}%
  \BibitemOpen
  \bibfield  {author} {\bibinfo {author} {\bibfnamefont {J.}~\bibnamefont
  {Fuchs}}, \bibinfo {author} {\bibfnamefont {C.~A.}\ \bibnamefont
  {Cecchetti}}, \bibinfo {author} {\bibfnamefont {M.}~\bibnamefont {Borghesi}},
  \bibinfo {author} {\bibfnamefont {T.}~\bibnamefont {Grismayer}}, \bibinfo
  {author} {\bibfnamefont {E.}~\bibnamefont {d'Humi{\`{e}}res}}, \bibinfo
  {author} {\bibfnamefont {P.}~\bibnamefont {Antici}}, \bibinfo {author}
  {\bibfnamefont {S.}~\bibnamefont {Atzeni}}, \bibinfo {author} {\bibfnamefont
  {P.}~\bibnamefont {Mora}}, \bibinfo {author} {\bibfnamefont {A.}~\bibnamefont
  {Pipahl}}, \bibinfo {author} {\bibfnamefont {L.}~\bibnamefont {Romagnani}},
  \bibinfo {author} {\bibfnamefont {A.}~\bibnamefont {Schiavi}}, \bibinfo
  {author} {\bibfnamefont {Y.}~\bibnamefont {Sentoku}}, \bibinfo {author}
  {\bibfnamefont {T.}~\bibnamefont {Toncian}}, \bibinfo {author} {\bibfnamefont
  {P.}~\bibnamefont {Audebert}}, \ and\ \bibinfo {author} {\bibfnamefont
  {O.}~\bibnamefont {Willi}},\ }\href {\doibase 10.1103/PhysRevLett.99.015002}
  {\bibfield  {journal} {\bibinfo  {journal} {Physical Review Letters}\
  }\textbf {\bibinfo {volume} {99}},\ \bibinfo {pages} {015002} (\bibinfo
  {year} {2007})}\BibitemShut {NoStop}%
\bibitem [{\citenamefont {Caprioli}\ \emph {et~al.}(2017)\citenamefont
  {Caprioli}, \citenamefont {Yi},\ and\ \citenamefont
  {Spitkovsky}}]{Caprioli2017}%
  \BibitemOpen
  \bibfield  {author} {\bibinfo {author} {\bibfnamefont {D.}~\bibnamefont
  {Caprioli}}, \bibinfo {author} {\bibfnamefont {D.~T.}\ \bibnamefont {Yi}}, \
  and\ \bibinfo {author} {\bibfnamefont {A.}~\bibnamefont {Spitkovsky}},\
  }\href {\doibase 10.1103/PhysRevLett.119.171101} {\bibfield  {journal}
  {\bibinfo  {journal} {Physical Review Letters}\ }\textbf {\bibinfo {volume}
  {119}},\ \bibinfo {pages} {171101} (\bibinfo {year} {2017})}\BibitemShut {NoStop}%
\bibitem [{\citenamefont {Caprioli}\ \emph {et~al.}(2015)\citenamefont
  {Caprioli}, \citenamefont {Pop},\ and\ \citenamefont
  {Spitkovsky}}]{Caprioli2015}%
  \BibitemOpen
 \bibfield  {author} {\bibinfo {author} {\bibfnamefont {D.}~\bibnamefont
  {Caprioli}}, \bibinfo {author} {\bibfnamefont {A.~R.}\ \bibnamefont {Pop}}, \
  and\ \bibinfo {author} {\bibfnamefont {A.}~\bibnamefont {Spitkovsky}},\
  }\href {\doibase 10.1088/2041-8205/798/2/L28} {\bibfield  {journal} {\bibinfo
   {journal} {Astrophysical Journal Letters}\ }\textbf {\bibinfo {volume}
  {798}},\ \bibinfo {pages} {L28} (\bibinfo {year} {2015})} \BibitemShut {NoStop}%
 \bibitem [{\citenamefont {Ohira}\ \emph {et~al.}(2016)\citenamefont {Ohira},
  \citenamefont {Kawanaka},\ and\ \citenamefont {Ioka}}]{Ohira2016a}%
  \BibitemOpen
 \bibfield  {author} {\bibinfo {author} {\bibfnamefont {Y.}~\bibnamefont
  {Ohira}}, \bibinfo {author} {\bibfnamefont {N.}~\bibnamefont {Kawanaka}}, \
  and\ \bibinfo {author} {\bibfnamefont {K.}~\bibnamefont {Ioka}},\ }\href
  {\doibase 10.1103/PhysRevD.93.083001} {\bibfield  {journal} {\bibinfo
  {journal} {Physical Review D}\ }\textbf {\bibinfo {volume} {93}},\ \bibinfo
  {pages} {083001} (\bibinfo {year} {2016})}\BibitemShut {NoStop}%
\end{thebibliography}

%

\end{document}